\documentclass[10pt,prd,
 amsmath,amssymb,
 aps, nofootinbib,twocolumn
]{revtex4-2}
\usepackage{graphicx}
\usepackage{tabularx}
\usepackage{dcolumn}
\usepackage{bm}
\usepackage{slashed}
\usepackage{setspace}
\usepackage[normalem]{ulem}
\usepackage{braket}
\usepackage{color}
\usepackage{subfigure}
\usepackage{float}
\usepackage{amsfonts}
\usepackage{booktabs}
\PassOptionsToPackage{hyphens}{url}
\usepackage[bookmarks=true,bookmarksopen=true,
bookmarksnumbered=true,
colorlinks=true,linkcolor=black,
citecolor=red]{hyperref}
\hypersetup{
    breaklinks=true
}
\definecolor{salmon}{RGB}{250,128,114} 
\usepackage{caption}

\begin{document}
\title{Fermion mass hierarchy and flavor signals from a $D_4$ dihedral symmetry within a 2HDM-Tx with four-zero textures}
\author{M. A. Arroyo-Ureña}
\email{marco.arroyo@fcfm.buap.mx}
\affiliation{Facultad de Ciencias Físico-Matemáticas and Centro Interdisciplinario de Investigación y Enseñanza de la Ciencia}


\author{Enrique D\'iaz}
\email{alcmaeon@gmail.com}
\affiliation{Sibatel Communications, 303 W Lincoln Ave $\#$140, Anaheim, CA, 92805.}

\author{J. L. Díaz-Cruz}
\email{jldiaz@fcfm.buap.mx}
\affiliation{Facultad de Ciencias Físico-Matemáticas and Centro Interdisciplinario de Investigación y Enseñanza de la Ciencia}

\author{H. Hernández-Arellano}
\email{col539008@colaborador.buap.mx}
\affiliation{Centro Interdisciplinario de Investigación y Enseñanza de la Ciencia}

\begin{abstract}
We propose an extended Texturized Two-Higgs Doublet Model (\text{2HDM-Tx}) where a discrete $D_4$ symmetry predicts a four-zero texture in the fermion mass matrices. In contrast to the standard ansatz where the underlying Yukawa matrices are assumed to be structurally identical, we introduce a novel configuration of non-equivalent matrices that complementarily generate a specific Hermitian four-zero mass structure. This framework naturally suppresses flavor-changing neutral currents and provides a structural rationale for the fermion mass hierarchy, while in addition including a dark matter candidate. As a concrete phenomenological application, we show that the model simultaneously accommodates the persistent neutral resonance excesses at $95\text{ GeV}$ and $152\text{ GeV}$, satisfies current \text{LHC} Higgs data, and is consistent with the stringent upper bounds from lepton flavor violation processes. These combined constraints severely restrict the parameter space of the scalar phenomenology. Finally, we study the prospects for the discovery of the lepton-flavor violating process $h \to \tau \mu$ at the High-Luminosity \text{LHC}.

\end{abstract}

\maketitle
\section{Introduction}
LHC data tells us that there is evidence for one SM-like Higgs particle~\cite{CMS:2012qbp, ATLAS:2012yve}, but theoretical motivations admit the possibility for a richer Higgs spectrum. Models with extra Higgs doublets have been studied in the literature, but it is not a trivial task to build realistic models that pass all experimental tests and add to our current understanding. In particular, the Yukawa sector of a general 2HDM has to be chosen with care, as otherwise it has problems with Flavor-Changing Neutral Currents (FCNC). In 2016, we published a paper titled ``Flavor violating Higgs signals in the Texturized Two-Higgs Doublet Model (2HDM-Tx)''~\cite{Arroyo-Urena:2013cyf}, in which we analyzed the general 2HDM with four-zero Yukawa matrices, such that when combined produce the so-called mass matrix with a four-zero texture. These matrices were constructed and motivated within a phenomenological framework. Specifically, we sought to answer the following question:
\begin{itemize}
    \item Is it possible to know the structure of the Yukawa matrices in such a way that it reproduces the observed values of the fermion masses, the hierarchy between them and the elements of the Cabbibo-Kobayashi-Maskawa matrix ($V_{\text{CKM}}$)~\cite{Kobayashi:1973fv}?
\end{itemize}
Bjorken \cite{PhysRevD.36.2109} showed that if the $(1,3)$ and $(3,1)$ entries are set to zero and the off-diagonal elements are assumed to be small compared with their diagonal counterparts in the mass matrices, then the relations among the elements of $V_{\text{CKM}}$ are correctly reproduced, and the values of $(V_{\text{CKM}})_{13}$ and $(V_{\text{CKM}})_{31}$ can be estimated to a good approximation, while the masses themselves emerge as theoretically well-defined eigenvalues of the mass matrix. Furthermore, to suppress dangerous FCNC mediated by neutral scalar bosons, it is customary to adopt Yukawa matrices with four-zero textures~\cite{Fritzsch:2002ga}, namely:

\begin{equation}
Y_i = \begin{pmatrix}\label{Yi}
0 & d_i & 0 \\
d^*_i & c_i & b_i \\
0 & b^*_i & a_i
\end{pmatrix},\quad (i=1,\,2).
\end{equation}
For 2HDM, given the form of the Yukawa matrix in Eq.~\eqref{Yi}, the mass matrix inherits the identical structure~\cite{Diaz-Cruz:2004wsi}:
\[
M^f = \begin{pmatrix}\label{MassMatrix}
0 & D_f & 0 \\
D^*_f & C_f & B_f \\
0 & B^*_f & A_f
\end{pmatrix}=\sum_{i=1}^2\frac{v_i}{\sqrt{2}}
\begin{pmatrix}
0 & d_i & 0 \\
d^*_i & c_i & b_i \\
0 & b^*_i & a_i
\end{pmatrix}.
\]

The four-zero texture is assumed for the Yukawa matrices in the gauge basis; hence, moving to the physical basis entails diagonalizing $M^f$, a step we address in due course.

It is therefore pertinent to pose the following question:
\begin{itemize}
    \item Can an underlying symmetry provide a natural justification for four-zero Yukawa matrices with distinct structural configurations?
\end{itemize}
The present work addresses this question by invoking the dihedral group $D_4$, thereby revealing its substantial predictive power. The dihedral group $  D_4  $ has already proved useful in distinct flavor-model constructions.  Early applications focused on the lepton sector, where $D_4$ (combined with an auxiliary $  Z_2  $ or $  Z_5  $) enforces maximal atmospheric mixing together with a vanishing reactor angle at leading order~\cite{Grimus:2004jr,Ishimori:2008fi}. Systematic analyzes of dihedral groups with preserved subgroups further demonstrate that $D_4$ can generate realistic quark-mass textures and relate the Cabibbo angle to pure group-theoretic quantities~\cite{Blum:2007jz}. Discrete flavor symmetries such as $D_4$ also arise naturally in intersecting D-brane constructions~\cite{Marchesano:2013xx}.

Thus, we aim to show that the 2HDM-Tx can simultaneously accommodate a wide range of experimental observables, including resonant excesses at 95 \cite{LEPWorkingGroupforHiggsbosonsearches:2003ing, CMS-PAS-HIG-20-002, ATLAS-CONF-2023-035, PhysRevD.109.035005, Ahriche:2023wkj} and 152~GeV \cite{Chen:2025vtg, Crivellin:2021ubm, Coloretti:2023wng, vonBuddenbrock:2017gvy, ATLAS:2021jbf, Bhattacharya:2023lmu}, LHC Higgs-boson measurements~\cite{ATLAS:2025qxq, CMS:2026nce}, stringent upper limits on lepton-flavor violating (LFV) processes~\cite{MEG:2016leq, BaBar:2009hkt, Belle:2021ysv}, as well as all relevant theoretical constraints. Furthermore, we show that the HL-LHC has the capability to test a distinctive prediction of our model, the flavor-violating decay $h\to\tau\mu$.

This work is structured as follows: In Sec.~\ref{sec:SecII}, we conduct a comprehensive analysis of the theoretical framework called the extended Texturized Two-Higgs Doublet Model (2HDM-Tx), with a particular emphasis on the theoretical implications of employing a four-zero texture in the Yukawa Lagrangian.  Experimental and theoretical constraints on the model parameter space are studied in Sec.~\ref{sec:SecIII}. Section~\ref{sec:SecIV} focuses on taking advantage of the insights gained from previous sections, performing a computational analysis of the proposed signal and its SM background processes. Finally, the conclusions are presented in Sec.~\ref{sec:SecV}.

\section{Theoretical framework}\label{sec:SecII}

The model contains two Higgs doublets $\Phi_{1,\,2}$, one flavon doublet $\eta$ (complex scalar), and the dark matter singlet $\chi$:
\begin{eqnarray*}
 \Phi_1&=&(\phi_1^+,\,\phi^0_1)^T,\;\Phi_2=(\phi_2^+,\,\phi^0_2)^T, \\ 
\eta &=& (\eta_1,\eta_2)^T,\
\chi.
\end{eqnarray*}

The assignments under the $SU(2)_L\times U(1)_Y$ group are given by 
\begin{eqnarray*}
 \Phi_{1,\,2} &\sim& (\textbf{2},\,+1/2), \\
 \eta&\sim&(\textbf{1},\,0),\\
 \chi&\sim&(\textbf{1},\,0).
\end{eqnarray*}
 A flavor-symmetry that exactly realizes the $2$ Yukawa splittings while automatically forbidding the three structural zeros (1,1), (1,3), (3,1) in both $Y_1$ and $Y_2$, can be built by invoking the dihedral group $D_4$ order 8 and non-abelian group, which admits the presentation
\[
D_4=\langle r,s\mid r^4=s^2=1,\; srs=r^{-1}\rangle,
\]
where \(r\) is the \(90^\circ\) rotation generator and \(s\) is a reflection generator.

The scalar fields transform under the $D_4$ symmetry in the following irreducible representations:
\begin{eqnarray}\label{transformations}
 \Phi_1&\sim& \mathbf{1}_{+ -},\ 
\Phi_2\sim \mathbf{1}_{- +}\nonumber \\ 
\eta &=& (\eta_1,\eta_2)^T\sim \mathbf{2},\
\chi \sim \mathbf{1}_{++}.
\end{eqnarray}

The explicit action of the generators is as follows:
\[
\begin{array}{r l}
& r\cdot \Phi_{1} = +\Phi_{1},\\
& r\cdot \Phi_{2} = -\Phi_{2},\\
& r\cdot \begin{pmatrix} \eta_{1} \\ \eta_{2} \end{pmatrix} = \begin{pmatrix} 0 & -1\\ 1 & 0 \end{pmatrix} \begin{pmatrix} \eta_{1} \\ \eta_{2} \end{pmatrix},\\
& r\cdot \chi = +\chi,
\end{array}
\
\begin{array}{r l}
& s\cdot \Phi_{1} = -\Phi_{1},\\
& s\cdot \Phi_{2} = +\Phi_{2},\\
& s\cdot \begin{pmatrix} \eta_{1} \\ \eta_{2} \end{pmatrix} = \begin{pmatrix} 1 & 0\\ 0 & -1 \end{pmatrix} \begin{pmatrix} \eta_{1} \\ \eta_{2} \end{pmatrix},\\
& s\cdot \chi = +\chi.
\end{array}
\]
The full group action on every element is obtained by successive application of \eqref{transformations}:
\begin{align*}
r^2:&\quad \Phi_1\to+\Phi_1,\quad\Phi_2\to+\Phi_2,\quad\eta\to-\eta, \\
r^3:&\quad \Phi_1\to+\Phi_1,\quad\Phi_2\to-\Phi_2,\quad\eta\to\begin{pmatrix}0&1\\-1&0\end{pmatrix}\eta, \\
rs:&\quad \Phi_1\to-\Phi_1,\quad\Phi_2\to-\Phi_2,\quad\eta\to\begin{pmatrix}0&1\\-1&0\end{pmatrix}\eta, \\
r^3s:&\quad \Phi_1\to-\Phi_1,\quad\Phi_2\to+\Phi_2,\quad\eta\to\begin{pmatrix}0&-1\\1&0\end{pmatrix}\eta,
\end{align*}
while $\chi$ transforms in a trivial way.

These rules are the unique faithful representation assignment (see Table \ref{tab:D4_fixed}) that automatically forbids the structural zeros at positions \((1,1)\), \((1,3)\) and \((3,1)\) in both Yukawa matrices while producing the four allowed parameters \(d,c,b,a\).
Unlike a 2HDM, the introduction of the flavon doublet $\eta\sim\mathbf{2}$ is essential in the $D_4$ model for three main reasons. 
\begin{itemize}
    \item First, it allows the generation of the four allowed Yukawa parameters $d,c,b,a$ in all 15 possible distributions between $\Phi_1$ and $\Phi_2$ (see Table~\ref{tab:D4_cases}) through higher-dimensional operators $\frac{y}{\Lambda}(\bar Q_L\eta)_{\rm inv}\Phi_2\psi_R$; without $\eta$, only a subset of textures would be realizable.
    \item Second, its vacuum expectation value provides a natural suppression scale $\Lambda$ that explains the smallness of first-generation masses, while the renormalizable terms account for the heavier fermions.
    \item Third, the $\kappa_1,\kappa_2$ interactions in the scalar potential mix the flavon components with the two Higgs doublets, leading to four CP-even scalars $S_k$ whose couplings to gauge bosons are modified by the factors $\xi_k = \cos\beta\,R_{k1}+\sin\beta\,R_{k2}$. This mixing is crucial to accommodate the LHC excesses at 95 and 152~GeV, and to produce deviations from the SM Higgs couplings that can be tested at future colliders. Hence, the flavon doublet is not an optional add-on but a necessary ingredient to realize the full phenomenological potential of the $D_4$ flavor symmetry.
\end{itemize}





\begin{table}[h]
\centering
\caption{Fixed representation assignments under the new \(D_4\) dihedral flavor symmetry model (one single model realizes every one of the 15 Yukawa splittings).}
\label{tab:D4_fixed}
\begin{tabular}{|c|c|}
\hline
\textbf{Field} & \textbf{Representation}  \\
\hline
\(Q_L^1\), \(u_R^1\) & \(\mathbf{1}_{++}\)  \\
\((Q_L^2, Q_L^3)\) & \(\mathbf{2}\)  \\
\((u_R^2, u_R^3)\) & \(\mathbf{2}\)  \\
\(\Phi_1\) & \(\mathbf{1}_{+-}\) \\
\(\Phi_2\) & \(\mathbf{1}_{-+}\)  \\
$\eta$ & \(\mathbf{2}\)  \\
$\chi$& \(\mathbf{1}_{++}\)\\
\hline
\end{tabular}
\end{table}

The \(D_4\) multiplication rules automatically forbid the structural zeros at positions (1,1), (1,3) and (3,1) in both \(Y_1\) and \(Y_2\).  
The four parameters \(d, c, b, a\) arise from the allowed singlet invariants of \(\mathbf{1}_{++} \times \mathbf{2}\) (for \(d\)) and \(\mathbf{2} \otimes \mathbf{2} \to \mathbf{1}_{+-} + \mathbf{1}_{-+}\) (for \(c, b, a\)) \footnote{For illustrative purposes, the invariance under these transformations for case 15 is presented in Appendix~\eqref{InvCases}.}.

\begin{table}[h]
\centering
\caption{Realization of all 15 Yukawa splittings under the same \(D_4\) representation assignment.}
\label{tab:D4_cases}
\begin{tabular}{c|l|l}
\hline
\textbf{Case} & \multicolumn{2}{c}{\textbf{Parameters generated by}} \\
 & \(\Phi_1\) (\(\mathbf{1}_{+-}\)) & \(\Phi_2\) (\(\mathbf{1}_{-+}\)) \\
\hline
2  & \(d, c\) & \(b, a\) \\
3  & \(d, b\) & \(c, a\) \\
4  & \(c, b\) & \(d, a\) \\
5  & \(b\) & \(d, c, a\) \\
6  & \(d\) & \(c, b, a\) \\
7  & \(c\) & \(d, b, a\) \\
8  & \(d, a\) & \(c, b\) \\
9  & \(b, a\) & \(d, c\) \\
10 & \(c, a\) & \(d, b\) \\
11 & \(d, c, a\) & \(b\) \\
12 & \(c, b, a\) & \(d\) \\
13 & \(d, b, a\) & \(c\) \\
14 & \(a\) & \(d, c, b\) \\
15 & \(d, c, b\) & \(a\) \\
\hline
\end{tabular}
\end{table}

The specific cases of the all Yukawa matrices that produce the four-zero mass matrix,
\begin{equation}\label{MassMatrix}
  M_f = \begin{pmatrix}
0 & D_f & 0 \\
D_f^* & C_f & B_f \\
0 & B_f^* & A_f
\end{pmatrix};\quad f=U,\,D,\ell, \nu,  
\end{equation}
are given as follows \footnote{We have made use of Grok 4.5 (xAI), a large language model, to identify candidate discrete groups; the remaining groups will be presented in future studies.}:
\\
 \\
\noindent
\begin{minipage}{0.48\textwidth}
\textbf{Case 1}
\[
Y_1 = \begin{pmatrix}
0 & d & 0 \\
d^* & c & b \\
0 & b^* & a
\end{pmatrix},\qquad
Y_2 = \begin{pmatrix}
0 & 0 & 0 \\
0 & 0 & 0 \\
0 & 0 & 0
\end{pmatrix}.
\]
\end{minipage}
\hfill
\begin{minipage}{0.48\textwidth}
\textbf{Case 2}
\[
Y_1 = \begin{pmatrix}
0 & d & 0 \\
d^* & c & 0 \\
0 & 0 & 0
\end{pmatrix},\qquad
Y_2 = \begin{pmatrix}
0 & 0 & 0 \\
0 & 0 & b \\
0 & b^* & a
\end{pmatrix}.
\]
\end{minipage}

\vspace{1cm}

\noindent
\begin{minipage}{0.48\textwidth}
\textbf{Case 3}
\[
Y_1 = \begin{pmatrix}
0 & d & 0 \\
d^* & 0 & b \\
0 & b^* & 0
\end{pmatrix},\qquad
Y_2 = \begin{pmatrix}
0 & 0 & 0 \\
0 & c & 0 \\
0 & 0 & a
\end{pmatrix}.
\]
\end{minipage}
\hfill
\begin{minipage}{0.48\textwidth}
\textbf{Case 4}
\[
Y_1 = \begin{pmatrix}
0 & 0 & 0 \\
0 & c & b \\
0 & b^* & 0
\end{pmatrix},\qquad
Y_2 = \begin{pmatrix}
0 & d & 0 \\
d^* & 0 & 0 \\
0 & 0 & a
\end{pmatrix}.
\]
\end{minipage}

\vspace{1cm}

\noindent
\begin{minipage}{0.48\textwidth}
\textbf{Case 5}
\[
Y_1 = \begin{pmatrix}
0 & 0 & 0 \\
0 & 0 & b \\
0 & b^* & 0
\end{pmatrix},\qquad
Y_2 = \begin{pmatrix}
0 & d & 0 \\
d^* & c & 0 \\
0 & 0 & a
\end{pmatrix}.
\]
\end{minipage}
\hfill
\begin{minipage}{0.48\textwidth}
\textbf{Case 6}
\[
Y_1 = \begin{pmatrix}
0 & d & 0 \\
d^* & 0 & 0 \\
0 & 0 & 0
\end{pmatrix},\qquad
Y_2 = \begin{pmatrix}
0 & 0 & 0 \\
0 & c & b \\
0 & b^* & a
\end{pmatrix}.
\]
\end{minipage}

\vspace{1cm}

\noindent
\begin{minipage}{0.48\textwidth}
\textbf{Case 7}
\[
Y_1 = \begin{pmatrix}
0 & 0 & 0 \\
0 & c & 0 \\
0 & 0 & 0
\end{pmatrix},\qquad
Y_2 = \begin{pmatrix}
0 & d & 0 \\
d^* & 0 & b \\
0 & b^* & a
\end{pmatrix}.
\]
\end{minipage}
\hfill
\begin{minipage}{0.48\textwidth}
\textbf{Case 8}
\[
Y_1 = \begin{pmatrix}
0 & d & 0 \\
d^* & 0 & 0 \\
0 & 0 & a
\end{pmatrix},\qquad
Y_2 = \begin{pmatrix}
0 & 0 & 0 \\
0 & c & b \\
0 & b^* & 0
\end{pmatrix}.
\]
\end{minipage}

\vspace{1cm}

\noindent
\begin{minipage}{0.48\textwidth}
\textbf{Case 9}
\[
Y_1 = \begin{pmatrix}
0 & 0 & 0 \\
0 & 0 & b \\
0 & b^* & a
\end{pmatrix},\qquad
Y_2 = \begin{pmatrix}
0 & d & 0 \\
d^* & c & 0 \\
0 & 0 & 0
\end{pmatrix}.
\]
\end{minipage}
\hfill
\begin{minipage}{0.48\textwidth}
\textbf{Case 10}
\[
Y_1 = \begin{pmatrix}
0 & 0 & 0 \\
0 & c & 0 \\
0 & 0 & a
\end{pmatrix},\qquad
Y_2 = \begin{pmatrix}
0 & d & 0 \\
d^* & 0 & b \\
0 & b^* & 0
\end{pmatrix}.
\]
\end{minipage}

\vspace{1cm}

\noindent
\begin{minipage}{0.48\textwidth}
\textbf{Case 11}
\[
Y_1 = \begin{pmatrix}
0 & d & 0 \\
d^* & c & 0 \\
0 & 0 & a
\end{pmatrix},\qquad
Y_2 = \begin{pmatrix}
0 & 0 & 0 \\
0 & 0 & b \\
0 & b^* & 0
\end{pmatrix}.
\]
\end{minipage}
\hfill
\begin{minipage}{0.48\textwidth}
\textbf{Case 12}
\[
Y_1 = \begin{pmatrix}
0 & 0 & 0 \\
0 & c & b \\
0 & b^* & a
\end{pmatrix},\qquad
Y_2 = \begin{pmatrix}
0 & d & 0 \\
d^* & 0 & 0 \\
0 & 0 & 0
\end{pmatrix}.
\]
\end{minipage}

\vspace{1cm}

\noindent
\begin{minipage}{0.48\textwidth}
\textbf{Case 13}
\[
Y_1 = \begin{pmatrix}
0 & d & 0 \\
d^* & 0 & b \\
0 & b^* & a
\end{pmatrix},\qquad
Y_2 = \begin{pmatrix}
0 & 0 & 0 \\
0 & c & 0 \\
0 & 0 & 0
\end{pmatrix}.
\]
\end{minipage}
\hfill
\begin{minipage}{0.48\textwidth}
\textbf{Case 14}
\[
Y_1 = \begin{pmatrix}
0 & 0 & 0 \\
0 & 0 & 0 \\
0 & 0 & a
\end{pmatrix},\qquad
Y_2 = \begin{pmatrix}
0 & d & 0 \\
d^* & c & b \\
0 & b^* & 0
\end{pmatrix}.
\]
\end{minipage}

\vspace{1cm}

\noindent
\begin{minipage}{0.48\textwidth}\label{caso15}
\textbf{Case 15}
\[
Y_1 = \begin{pmatrix}
0 & d & 0 \\
d^* & c & b \\
0 & b^* & 0
\end{pmatrix},\qquad
Y_2 = \begin{pmatrix}
0 & 0 & 0 \\
0 & 0 & 0 \\
0 & 0 & a
\end{pmatrix}.
\]
\end{minipage}

\subsection{Scalar Potential}



The full renormalizable scalar potential invariant under \(D_4\) is \footnote{Invariance under $D_4$ symmetry of the scalar potential is presented in Appendix~\eqref{InvPotential}.}
\begin{align}\label{ScalarPotential}
V_{D_4} &= m_{11}^2 |\Phi_1|^2 + m_{22}^2 |\Phi_2|^2 
+ \frac{\lambda_1}{2}(|\Phi_1|^2)^2 + \frac{\lambda_2}{2}(|\Phi_2|^2)^2\nonumber\\ 
&+ \lambda_3 |\Phi_1|^2|\Phi_2|^2 + \lambda_4 |\Phi_1^\dagger\Phi_2|^2 \nonumber 
+ m_\eta^2 |\eta|^2 + \lambda_\eta (|\eta|^2)^2 \\
&+ \lambda_{\Phi\eta} \bigl(|\Phi_1|^2 + |\Phi_2|^2\bigr)|\eta|^2 
+ \kappa_1 \bigl(\Phi_1^\dagger \Phi_2 \eta_1\eta_2 + \text{h.c.}\bigr)\nonumber\\
&+ \kappa_2 \bigl(\Phi_1^\dagger \Phi_1 \eta_1^2 + \Phi_2^\dagger \Phi_2 \eta_2^2 + \text{h.c.}\bigr)+ V_{\text{DM}}, 
\end{align}
where the dark matter sector is given by

\begin{eqnarray}
V_{\text{DM}} &=& \frac12 m_{\chi}^2 \chi^2 + \frac{\lambda_\chi}{4!} \chi^4 
+ \frac{\lambda_{\chi H}}{4} \chi^2 \left(|\Phi_1|^2 + |\Phi_2|^2\right)\nonumber\\
&+& \frac{\lambda_{\chi \eta}}{4} \chi^2 |\eta|^2.
\end{eqnarray}

A detailed analysis of the viability of the dark matter candidate is beyond the scope of the present work. In a future study, we plan to investigate its phenomenological implications within the parameter space identified further below, including the computation of the dark matter relic abundance and its compatibility with the observed cosmological
relic density, as well as the constraints from direct detection
experiments. This future analysis will further restrict the viable parameter space of the model and explore the interplay between the dark matter and scalar sectors.

\subsubsection{Determination of Quadratic Coefficients and CP-Even Mass Matrix from the Full UV Scalar Potential}

To determine the quadratic mass parameters \(m_{11}^2\), \(m_{22}^2\), and \(m_\eta^2\) and the CP-even mass matrix, we expand the neutral components around their vacuum expectation values (VEV), assuming CP conservation and real aligned VEVs:
\begin{eqnarray}
    \Phi_1^0 &=& \frac{v_1 + h_1}{\sqrt{2}}, \quad
\Phi_2^0 = \frac{v_2 + h_2}{\sqrt{2}},\nonumber\\
\eta_1 &=& \frac{v_{\eta1} + h_{\eta1}}{\sqrt{2}}, \quad
\eta_2 = \frac{v_{\eta2} + h_{\eta2}}{\sqrt{2}}.\nonumber
\end{eqnarray}

The minimum condition requires that the first derivatives of $V_{D_4}$ with respect to the CP-even fluctuation fields vanish at the VEVs. This produces the following four exact tadpole equations:


\begin{equation}
\begin{aligned}
\frac{\partial V}{\partial h_1}\bigg|_{\mathrm{VEVs}}
= &\, m_{11}^2 v_1 + \frac{1}{2}\lambda_1 v_1^3 + \frac{1}{2}(\lambda_3+\lambda_4)v_1 v_2^2 \\
&+ \frac{1}{2}\lambda_{\Phi\eta} v_1 (v_{\eta 1}^2+v_{\eta 2}^2) + \frac{1}{2}\kappa_1 v_2 v_{\eta 1} v_{\eta 2} \\
&+ \kappa_2 v_1 v_{\eta 1}^2 = 0.
\end{aligned}\label{eq:tad1}
\end{equation}

\begin{equation}
\begin{aligned}
\frac{\partial V}{\partial h_2}\bigg|_{\mathrm{VEVs}}
= &\, m_{22}^2 v_2 + \frac{1}{2}\lambda_2 v_2^3 + \frac{1}{2}(\lambda_3+\lambda_4)v_1^2 v_2 \\
&+ \frac{1}{2}\lambda_{\Phi\eta} v_2 (v_{\eta 1}^2+v_{\eta 2}^2) + \frac{1}{2}\kappa_1 v_1 v_{\eta 1} v_{\eta 2} \\
&+ \kappa_2 v_2 v_{\eta 2}^2 = 0.
\end{aligned}\label{eq:tad2}
\end{equation}

\begin{equation}
\begin{aligned}
\frac{\partial V}{\partial h_{\eta 1}}\bigg|_{\mathrm{VEVs}}
= &\, m_\eta^2 v_{\eta 1} + \lambda_\eta v_{\eta 1}(v_{\eta 1}^2+v_{\eta 2}^2) \\
&+ \frac{1}{2}\lambda_{\Phi\eta}(v_1^2+v_2^2)v_{\eta 1} + \frac{1}{2}\kappa_1 v_1 v_2 v_{\eta 2} \\
&+ \kappa_2 v_1^2 v_{\eta 1} = 0.
\end{aligned}\label{eq:tad3}
\end{equation}

\begin{equation}
\begin{aligned}
\frac{\partial V}{\partial h_{\eta 2}}\bigg|_{\mathrm{VEVs}}
= &\, m_\eta^2 v_{\eta 2} + \lambda_\eta v_{\eta 2}(v_{\eta 1}^2+v_{\eta 2}^2) \\
&+ \frac{1}{2}\lambda_{\Phi\eta}(v_1^2+v_2^2)v_{\eta 2} + \frac{1}{2}\kappa_1 v_1 v_2 v_{\eta 1}\\
&+ \kappa_2 v_2^2 v_{\eta 2} = 0.
\end{aligned}\label{eq:tad4}
\end{equation}

Solving the first two tadpole equations (\ref{eq:tad1}) and (\ref{eq:tad2}) for the Higgs quadratic parameters gives the exact expressions for \(m_{11}^2\) and \(m_{22}^2\):


\begin{align}
m_{11}^2 &= -\frac{1}{2}\lambda_1 v_1^2 - \frac{1}{2}(\lambda_3+\lambda_4)v_2^2 - \frac{1}{2}\lambda_{\Phi\eta}(v_{\eta 1}^2+v_{\eta 2}^2) \nonumber\\
&- \frac{\kappa_1 v_2 v_{\eta 1} v_{\eta 2}}{2 v_1} - \kappa_2 v_{\eta 1}^2, \label{eq:m11full}\\
m_{22}^2 &= -\frac{1}{2}\lambda_2 v_2^2 - \frac{1}{2}(\lambda_3+\lambda_4)v_1^2 - \frac{1}{2}\lambda_{\Phi\eta}(v_{\eta 1}^2+v_{\eta 2}^2)\nonumber \\
&- \frac{\kappa_1 v_1 v_{\eta 1} v_{\eta 2}}{2 v_2} - \kappa_2 v_{\eta 2}^2.\label{eq:m22full}
\end{align}
The solution of the last two tadpole equations, (\ref{eq:tad3}) and (\ref{eq:tad4}), yield consistent expressions given by


\begin{align}
m_\eta^2 &= -\lambda_\eta (v_{\eta 1}^2+v_{\eta 2}^2) - \frac{1}{2}\lambda_{\Phi\eta}(v_1^2+v_2^2) \nonumber\\
&- \frac{\kappa_1 v_1 v_2 v_{\eta 2}}{2 v_{\eta 1}} - \kappa_2 v_1^2, \\
m_\eta^2 &= -\lambda_\eta (v_{\eta 1}^2+v_{\eta 2}^2) - \frac{1}{2}\lambda_{\Phi\eta}(v_1^2+v_2^2)\nonumber\\
&- \frac{\kappa_1 v_1 v_2 v_{\eta 1}}{2 v_{\eta 2}} - \kappa_2 v_2^2.\label{eq:meta}
\end{align}

These are the exact quadratic coefficients expressed in terms of the quartic couplings and the VEVs.

To relate the quartic couplings to the physical scalar masses, we now construct the full 4×4 CP-even mass-squared matrix. This matrix is obtained by taking the second derivatives of the full potential \(V_{D_4}\) with respect to the CP-even fluctuation fields \(h_1, h_2, h_{\eta1}\) and \(h_{\eta2}\) and evaluating at the VEVs. The resulting symmetric matrix \(\mathcal{M}^2\) has the following elements:
\[
\mathcal{M}^2_{11} = 2\lambda_1 v_1^2 + (\lambda_3 + \lambda_4) v_2^2 + \lambda_{\Phi\eta} (v_{\eta1}^2 + v_{\eta2}^2) + 2\kappa_2 v_{\eta1}^2,
\]
\[
\mathcal{M}^2_{22} = 2\lambda_2 v_2^2 + (\lambda_3 + \lambda_4) v_1^2 + \lambda_{\Phi\eta} (v_{\eta1}^2 + v_{\eta2}^2) + 2\kappa_2 v_{\eta2}^2,
\]
\[
\mathcal{M}^2_{33} = 2\lambda_\eta v_{\eta1}^2 + \lambda_\eta v_{\eta2}^2 + \lambda_{\Phi\eta} (v_1^2 + v_2^2) + 2\kappa_2 v_1^2,
\]
\[
\mathcal{M}^2_{44} = 2\lambda_\eta v_{\eta2}^2 + \lambda_\eta v_{\eta1}^2 + \lambda_{\Phi\eta} (v_1^2 + v_2^2) + 2\kappa_2 v_2^2,
\]
\[
\mathcal{M}^2_{12} = \mathcal{M}^2_{21} = (\lambda_3 + \lambda_4) v_1 v_2 + \kappa_1 v_{\eta1} v_{\eta2},
\]
\[
\mathcal{M}^2_{13} = \mathcal{M}^2_{31} = \lambda_{\Phi\eta} v_1 v_{\eta1} + 2\kappa_2 v_1 v_{\eta1},
\]
\[
\mathcal{M}^2_{14} = \mathcal{M}^2_{41} = \lambda_{\Phi\eta} v_1 v_{\eta2} + \kappa_1 v_2 v_{\eta2},
\]
\[
\mathcal{M}^2_{23} = \mathcal{M}^2_{32} = \lambda_{\Phi\eta} v_2 v_{\eta1} + \kappa_1 v_1 v_{\eta1},
\]
\[
\mathcal{M}^2_{24} = \mathcal{M}^2_{42} = \lambda_{\Phi\eta} v_2 v_{\eta2} + 2\kappa_2 v_2 v_{\eta2},
\]
\[
\mathcal{M}^2_{34} = \mathcal{M}^2_{43} = \lambda_\eta v_{\eta1} v_{\eta2} + \kappa_1 v_1 v_2.
\]


The four physical CP-even scalar masses (two Higgs bosons + two flavon scalars) are the eigenvalues of this 4×4 matrix \(\mathcal{M}^2\). In a phenomenological study, one chooses the quartic couplings \(\lambda_1, \lambda_2, \lambda_3, \lambda_4, \lambda_\eta, \lambda_{\Phi\eta}\) (and \(\kappa_1, \kappa_2\) if needed) such that the eigenvalues of \(\mathcal{M}^2\) exactly match the desired physical masses. The quadratic parameters \(m_{11}^2\), \(m_{22}^2\), and \(m_\eta^2\) are then fixed by the expressions \eqref{eq:m11full} and \eqref{eq:meta} in terms of these quartics and the VEVs.

\subsubsection{Explicit masses \( M_{S_k} \) from the CP-even mass matrix}

The $4\times4$ CP-even mass-squared matrix \(\mathcal{M}^2\) is diagonalized by a real orthogonal 4×4 mixing matrix $R$ (i.e. $R^T R$ = $\textbf{1}$) via
\[
R^T \mathcal{M}^2 R = \operatorname{diag}(m_{S_1}^2,\, m_{S_2}^2,\, m_{S_3}^2,\, m_{S_4}^2).
\]

The physical CP-even scalar masses are:
\[
M_{S_k} = \sqrt{\rho_k}, \qquad k=1,2,3,4,
\]
where \(\rho_k\) (\(k=1,2,3,4\)) are the four eigenvalues of the explicit matrix \(\mathcal{M}^2\). In practice, for any chosen set of quartic couplings \(\lambda_i\), \(\kappa_i\) and VEVs, the four masses \(m_{S_k}\) and the orthogonal matrix \(R\) are obtained by numerical diagonalization of \(\mathcal{M}^2\).

They satisfy the characteristic equation:

\[
\det(\mathcal{M}^2 - m^2 \mathbb{I}) = 0,
\]

which expands to:

\[
m^8 - (\operatorname{tr} \mathcal{M}^2) \, m^6 + \Sigma_2 \, m^4 - \Sigma_3 \, m^2 + \det \mathcal{M}^2 = 0,
\]

where:

\[
\begin{aligned}
\operatorname{tr} \mathcal{M}^2 &= \mathcal{M}^2_{11} + \mathcal{M}^2_{22} + \mathcal{M}^2_{33} + \mathcal{M}^2_{44}, \\
\Sigma_2 &= \sum_{i<j} \left( \mathcal{M}^2_{ii} \mathcal{M}^2_{jj} - (\mathcal{M}^2_{ij})^2 \right), \\
\Sigma_3 &= \sum_{i<j<k} \det
\begin{pmatrix}
\mathcal{M}^2_{ii} & \mathcal{M}^2_{ij} & \mathcal{M}^2_{ik} \\
\mathcal{M}^2_{ji} & \mathcal{M}^2_{jj} & \mathcal{M}^2_{jk} \\
\mathcal{M}^2_{ki} & \mathcal{M}^2_{kj} & \mathcal{M}^2_{kk}
\end{pmatrix}.
\end{aligned}
\]

The four solutions \( m^2 = m_{S_1}^2, m_{S_2}^2, m_{S_3}^2, m_{S_4}^2 \) are the eigenvalues of \( \mathcal{M}^2 \). In general, they are the roots of a quartic equation:



\[
m^8 - \omega_3 \, m^6 + \omega_2 \, m^4 - \omega_1 \, m^2 + \omega_0 = 0,
\]

with \( \omega_3 = \operatorname{tr} \mathcal{M}^2 \), \( \omega_2 = \Sigma_2 \), \( \omega_1 = \Sigma_3 \), \( \omega_0 = \det \mathcal{M}^2 \). The explicit expressions in terms of the parameters \( \lambda_i, \kappa_i, v_i, v_{\eta i} \) are given by

\begin{eqnarray}
 M_{S_1}^2&=&\frac{1}{2}\left( \mathcal{M}^2_{11}+\mathcal{M}^2_{22}
 + \sqrt{(\mathcal{M}^2_{11}-\mathcal{M}^2_{22})^2 + 4 (\mathcal{M}^2_{12})^2} \right), \nonumber\\
 M_{S_2}^2 &=& \frac{1}{2}\left( \mathcal{M}^2_{11}+\mathcal{M}^2_{22} - \sqrt{(\mathcal{M}^2_{11}-\mathcal{M}^2_{22})^2 + 4 (\mathcal{M}^2_{12})^2} \right), \nonumber\\
M_{S_3}^2 &=& \frac{1}{2}\left( \mathcal{M}^2_{33}+\mathcal{M}^2_{44} + \sqrt{(\mathcal{M}^2_{33}-\mathcal{M}^2_{44})^2 + 4 (\mathcal{M}^2_{34})^2} \right), \nonumber\\
M_{S_4}^2 &=& \frac{1}{2}\left( \mathcal{M}^2_{33}+\mathcal{M}^2_{44} -\sqrt{(\mathcal{M}^2_{33}-\mathcal{M}^2_{44})^2 + 4 (\mathcal{M}^2_{34})^2} \right). \nonumber\\
\end{eqnarray}

In a similar way, we obtained the CP-odd masses where their analytical expressions are given as follows:
\begin{eqnarray}
    M^2_{A_1}&=& \frac{2\kappa_1v_{\eta1}v_{\eta2}}{\sin(2\beta)},\nonumber\\
    M^2_{A_2}&=& 4\lambda_\eta v^2_{\eta1}+\lambda_{\Phi\eta}v^2+\kappa_1 v_1 v_2 ,\nonumber\\
M^2_{A_3}&=& 2\lambda_\eta v^2_{\eta2}+\lambda_{\Phi\eta}v^2-\kappa_1 v_1 v_2. \nonumber
\end{eqnarray}
Finally, the charged scalar mass is given by
\begin{equation}
    M_{H^\pm}^2=M_{A_1}^2-\frac{\lambda_4}{2}v^2.
\end{equation}
\subsection{Yukawa Lagrangian for the \(D_4\) Model}



The complete \(D_4\)-invariant Yukawa Lagrangian (before the flavon acquires a vacuum expectation value) is
\[
\mathcal{L}_Y = \mathcal{L}_Y^U + \mathcal{L}_Y^D + \mathcal{L}_Y^\ell + \mathcal{L}_Y^\nu + \text{h.c.},
\]

where the individual sectors are given as follows

\begin{eqnarray}
\mathcal{L}_Y^U &=& y_d^U\,\bar{Q}_L^2\,\tilde{\Phi}_1\,u_R^1 + y_c^U\,\bar{Q}_L^2\,\tilde{\Phi}_2\,u_R^2 + y_b^U\,\bar{Q}_L^2\,\tilde{\Phi}_2\,u_R^3\nonumber\\
&+&
y_a^U\,\bar{Q}_L^3\,\tilde{\Phi}_2\,u_R^3 + \frac{y_\eta^U}{\Lambda}\big(Q_L\eta \big)_{\mathrm{inv}}\tilde{\Phi}_2\,u_R, \\
\mathcal{L}_Y^D &=& y_d^D\,\bar{Q}_L^2\,\Phi_1\,d_R^1 + y_c^D\,\bar{Q}_L^2\,\Phi_2\,d_R^2 + y_b^D\,\bar{Q}_L^2\,\Phi_2\,d_R^3\nonumber\\
&+&
y_a^D\,\bar{Q}_L^3\,\Phi_2\,d_R^3 + \frac{y_\eta^D}{\Lambda}\big(Q_L\eta \big)_{\mathrm{inv}}\Phi_2\,d_R, \\[0.5em]
\mathcal{L}_Y^\ell &=& y_a^\ell\,\bar{L}_L^2\,\Phi_1\,\ell_R^1 + y_c^\ell\,\bar{L}_L^2\,\Phi_2\,\ell_R^2 + y_b^\ell\,\bar{L}_L^2\,\Phi_2\,\ell_R^3\nonumber\\
&+&
y_a^\ell\,\bar{L}_L^3\,\Phi_2\,\ell_R^3 + \frac{y_\eta^\ell}{\Lambda}\big(\bar{L}_L\eta \big)_{\mathrm{inv}}\Phi_2\,\ell_R, \\
\mathcal{L}_Y^\nu &=& y_a^\nu\,\bar{L}_L^2\,\tilde{\Phi}_1\,\nu_R^1 + y_c^\nu\,\bar{L}_L^2\,\tilde{\Phi}_2\,\nu_R^2 + y_b^\nu\,\bar{L}_L^2\,\tilde{\Phi}_2\,\nu_R^3\nonumber\\
&+&
y_a^\nu\,\bar{L}_L^3\,\tilde{\Phi}_2\,\nu_R^3 + \frac{y_\eta^\nu}{\Lambda}\big(\bar{L}_L\eta \big)_{\mathrm{inv}}\tilde{\Phi}_2\,\nu_R,    
\end{eqnarray}
where every bilinear \(\bar{\Psi}_L \Phi_k \psi_R\) stands for the standard SU(2)\(_L\)-invariant contraction
\[
\bar{\Psi}_{L\alpha} \, (i\sigma_2 \Phi_k^*)^\alpha \, \psi_R.
\] We have denoted $U=u,\,c,\,t$, $D=d,\,s,\,b$, $\ell=e,\,\mu,\,\tau$ and $\nu=\nu_\mu,\,\nu_e,\,\nu_\tau$. 

The term $\bigl(\overline{\Psi}_L^{\,f} \eta\bigr)_{\text{inv}}$ denotes the unique $D_4$-singlet contraction of the product of two doublets ($\mathbf{2}\otimes\mathbf{2} \to \mathbf{1}_{++}$). After $\eta$ acquires a VEV $\langle\eta\rangle = (v_{\eta1}, v_{\eta2})^T$, this dimension-5 operator (suppressed by the scale $\Lambda$) generates the remaining allowed matrix entries, specifically the off-diagonal elements that involve the first generation, explaining the fermion mass hierarchy.

\noindent


In the $D_4$ model, no renormalizable Yukawa coupling directly involves the first-generation fermions because their $\mathbf{1}_{++}$ representation does not combine with the Higgs doublets ($\mathbf{1}_{+-}$ or $\mathbf{1}_{-+}$) to form a $D_4$ singlet. Consequently, the bare mass matrix contains zeros in the $(1,1)$, $(1,3)$ and $(3,1)$ entries. Nevertheless, after electroweak symmetry breaking, the mass matrix is diagonalized by unitary transformations that mix all three generations. This mixing induces effective Higgs couplings to the first-generation fermions even though they are absent from the original Lagrangian: the physical light fermion $f_1$ acquires a component of the heavier doublet fields $Q_L^2$ and $Q_L^3$, which couple to the Higgs. Therefore, the interactions of the Higgs with the first generation arise entirely from fermion mixing and the non‑zero off‑diagonal entries of the Yukawa matrices, providing a natural explanation for the smallness of the first‑generation masses and their Higgs couplings.

After the flavon doublet \(\eta\) acquires a vacuum expectation value \(\langle\eta\rangle = (v_{\eta1}, v_{\eta2})^T \neq 0\), the higher-dimensional operators produce effective renormalizable contributions to the remaining allowed matrix entries in every sector. The \(D_4\) multiplication rules automatically forbid the three structural zeros at positions (1,1), (1,3) and (3,1) in all Yukawa matrices while allowing the four parameters \(d,c,b,a\) (with independent coefficients in each sector) to be assigned to \(\Phi_1\) or \(\Phi_2\) in any of the 15 ways listed in Table~\ref{tab:D4_cases}.



\subsection{Neutral Yukawa Lagrangian}

After electroweak symmetry breaking and insertion of the flavon vacuum expectation value \(\langle\eta\rangle = (v_{\eta1},v_{\eta2})^T\), the neutral fields are expanded as follows:
\begin{align}
   \phi_1^0&=\frac{1}{\sqrt{2}}(v_1+h_1+i a_1),\\
    \phi_2^0&=\frac{1}{\sqrt{2}}(v_2+h_2+i a_2),\\ 
    \eta_1^0&=\frac{1}{\sqrt{2}}(v_{\eta1}+h_{\eta1}+i a_{\eta1}),\\
    \eta_2^0&=\frac{1}{\sqrt{2}}(v_{\eta2}+h_{\eta2}+i a_{\eta2}).
\end{align}

The interaction part of the Yukawa Lagrangian in the flavor basis (before going to the physical fermion mass eigenstates) is
\begin{equation}
 \mathcal{L}_{\rm int}^f = \frac{1}{\sqrt{2}} \bar{\Psi}_L^f \bigl( h_1 Y_1^f + h_2 Y_2^f +h_{\eta1}Y_{\eta1}^f+h_{\eta2}Y_{\eta2}^f\bigr) \psi_R^f + \text{h.c.}   \end{equation}

The full set of CP-even interaction eigenstates, $\textbf{h}$,  are collected in the vectors
\[
\mathbf{h} = \begin{pmatrix} h_1 \\ h_2 \\ h_{\eta1} \\ h_{\eta2} \end{pmatrix}.
\]

The corresponding physical mass eigenstates \(S_k\) are the explicit linear combinations
\begin{align}
S_1 &= R_{11} h_1 + R_{12} h_2 + R_{13} h_{\eta1} + R_{14} h_{\eta2}, \nonumber \\
S_2 &= R_{21} h_1 + R_{22} h_2 + R_{23} h_{\eta1} + R_{24} h_{\eta2}, \nonumber \\
S_3 &= R_{31} h_1 + R_{32} h_2 + R_{33} h_{\eta1} + R_{34} h_{\eta2}, \nonumber \\
S_4 &= R_{41} h_1 + R_{42} h_2 + R_{43} h_{\eta1} + R_{44} h_{\eta2}.
\end{align}

The corresponding analysis for the imaginary part of $\phi^0$ is done similarly to the real part.


\subsection*{Neutral Yukawa Lagrangian for all fermion sectors}

Let $f_i = (U, D, \ell, \nu)_i$ denote the four fermion sectors. For each sector, the physical fermions are $f_i$ ($i=1,2,3$) with masses $m_i^f$, and the rotated Yukawa matrices are defined as
\begin{equation}\label{rotated_Yukawas}
\tilde Y_{1,ij}^f = (U_L^{f\dagger} Y_1^f U_R^f)_{ij}, \qquad
\tilde Y_{2,ij}^f = (U_L^{f\dagger} Y_2^f U_R^f)_{ij},    
\end{equation}

where $U_{L,R}^f$ diagonalize the mass matrix $M^f = (v_1 Y_1^f + v_2 Y_2^f)/\sqrt{2}$. Under the assumption that $M^f$ is hermitian, it follows that $U_{L}^f=U_{R}^f\equiv U^f$. Here, $U^f=\mathcal{O}^fP^f$, where $P^f=\text{diag}(e^{i\alpha_1^f},\,e^{i\alpha_2^f},\,1)$ and $\mathcal{O}^f$ is given by Eq.~\eqref{MatrixO}. 
\begin{figure*}[t] 
\centering
\footnotesize 
\setlength{\arraycolsep}{3pt} 
\begin{equation} \label{MatrixO}
\mathcal{O}_{f}=\left(\begin{array}{ccc}
    \sqrt{\frac{m_{f_2}m_{f_3}(A-m_{f_1})}{A(m_{f_2}-m_{f_1})(m_{f_3}-m_{f_1})}} & \sqrt{\frac{m_{f_1}m_{f_3}(m_{f_2}-A)}{A(m_{f_2}-m_{f_1})(m_{f_3}-m_{f_2})}} & \sqrt{\frac{m_{f_1}m_{f_3}(A-m_{f_3})}{A(m_{f_3}-m_{f_1})(m_{f_3}-m_{f_2})}}\\ \\ \vspace{3mm}
    -\sqrt{\frac{m_{f_1}(m_{f_1}-A)}{(m_{f_2}-m_{f_1})(m_{f_3}-m_{f_1})}} & \sqrt{\frac{m_{f_2}(A-m_{f_2})}{(m_{f_2}-m_{f_1})(m_{f_3}-m_{f_2})}} & \sqrt{\frac{m_{f_3}(m_{f_2}-A)}{(m_{f_2}-m_{f_1})(m_{f_3}-m_{f_2})}}\\ \vspace{3mm}
    \sqrt{\frac{m_{f_1}(A-m_{f_2})(A-m_{f_3})}{A(m_{f_2}-m_{f_1})(m_{f_3}-m_{f_1})}} & -\sqrt{\frac{m_{f_2}(A-m_{f_1})(m_{f_3}-A)}{A(m_{f_2}-m_{f_1})(m_{f_3}-m_{f_2})}} & \sqrt{\frac{m_{f_3}(A-m_{f_1})(A-m_{f_2})}{A(m_{f_3}-m_{f_1})(m_{f_3}-m_{f_2})}}
\end{array}\right),
\end{equation}
\\
\text{where $m_{f_i}$ ($i=1,\,2,\,3$) are the physical fermion masses of the three generations.}
\end{figure*}

The elements of the mass matrix $M_f$ in Eq.~\eqref{MassMatrix} are related to the fermion masses $m_f$ through the
principal invariants:


\begin{eqnarray}\label{invariants}
I_1 = \text{Tr}(M_f)&=&C_f+A_f=m_{f_1}+m_{f_2}+m_{f_3},\nonumber\\
I_2&=&C_fA_f-D_f^2-B_f^{2}\nonumber\\
&=&m_{f_1}m_{f_2}+m_{f_1}m_{f_3}+m_{f_2}m_{f_3}\\
I_3=\text{det}(M_f)&=&-D_f^2A_f=m_{f_1}m_{f_2}m_{f_3}\nonumber.
\end{eqnarray}

From Eq.~\eqref{invariants}, we find a relation between the components of the mass matrix of four-zero textures and the fermion masses as follows:
\begin{eqnarray}\label{MassMatrixElementA}
    A_f&=&m_{f_3}(1-r_2\gamma_f),\nonumber\\
    B_f&=&m_{f_3}\sqrt{\frac{r_2(r_2\gamma_f+r_1-1)(r_2\gamma_f-1)}{1-r_2\gamma_f}},\\
    C_f&=&m_{f_3}(r_2\gamma_f+r_1+r_2),\nonumber\\
    D_f&=&\sqrt{-\frac{m_{f_1}m_{f_2}}{1-r_2\gamma_f}}\nonumber,
\end{eqnarray}
where $r_i=m_{f_i}/m_{f_3}$, and we parameterize $\gamma_f\in(0,\,1)$.

The CP-even scalars $S_i$ and CP-odd scalars $A_j$ ($i=1,\,2,\,3,\,4;\,j=2,\,3, \,4$) are related to the interaction eigenstates by $h_i = \sum_k R_{ki} S_k$ and $a_i = \sum_k R^A_{ki} A_k$ (with $h_3,h_4$ and $a_3,a_4$ coming from the flavon doublet).



Using the mass relation 
\begin{equation}
 \tilde Y_{1,ij}^f = \frac{\sqrt{2}}{v_1} \bar{M}^f \delta_{ij} - \frac{v_2}{v_1} \tilde Y_{2,ij}^f,   
\end{equation}

the neutral Yukawa Lagrangian in terms of mass eigenstates takes the following explicit form:

\begin{equation}\label{FullYuwLag}
\begin{aligned}
\mathcal{L}_\text{Y}^0 
&= -  S_k  \bar{f}_i 
\left[
\frac{R_{k1}^S}{v_1} \bar{M}^f \delta_{ij} 
+ \frac{\Delta R_k^S}{\sqrt{2}} \Big(\tilde{Y}_{2,ij}^f 
+  \mathcal{F}_{ij}^f\Big)
\right] f_j \\
&+ i\epsilon_f  A_k \bar{f}_i 
\left[
 \frac{R_{k1}^A}{v_1} \bar{M}^f \delta_{ij} 
+ \frac{\Delta R_k^A}{\sqrt{2}} \Big(\tilde{Y}_{2,ij}^f 
+  \mathcal{F}_{ij}^f\Big)
\right] \gamma^5 f_j \\
&+\text{h.c.}
\end{aligned}
\end{equation}
We define the auxiliary combinations of the scalar mixing matrices:
\[
\Delta R_k^{S,A} \equiv R_{k2}^{S,A} - \frac{v_2}{v_1} R_{k1}^{S,A},
\]
and the flavon-induced effective Yukawa matrix:
\begin{equation}\label{Supresion_Lambda}
\mathcal{F}_{ij}^f \equiv \frac{1}{2\Lambda} \bigl( v_{\eta_ 1}\,\delta_{i2} + v_{\eta_2}\,\delta_{i3} \bigr) \,\tilde Y_{2,j}^f,    
\end{equation}

where \(\tilde Y_{2,j}^f\) are the rotated flavon couplings to the right-handed fermions \(f_R^j\), and 
\[
\epsilon_f = 
\begin{cases}
+1, & f = u, \nu, \\
-1, & f = d, \ell.
\end{cases}
\]
 The term proportional to \(\bar{M}^f \delta_{ij}\) in Eq.~\eqref{FullYuwLag} gives the usual SM-like Higgs couplings when the alignment limit is taken (\(R_{11}=v_1/v\), \(R^A_{11}=0\), etc.). The terms involving \(\tilde Y_{2,ij}^{f}\) arise from the second Higgs doublet \(\Phi_2\). The case for \(i\neq j\), induces tree-level FCNCs in the scalar sector and the terms involving \(\mathcal{F}_{ij}^f\) arise from the dimension-5 flavon operator. They are suppressed by \(\Lambda\) and carry the characteristic flavor structure \(\delta_{i2}, \delta_{i3}\) from the flavon VEV, explaining the smallness of the first-generation couplings. The flavon fluctuations do not appear explicitly because they have been rotated into the physical scalars \(S_k\) and \(A_k\) via the mixing matrices \(R\) and \(R^A\), which are encoded in the factors \(R_{k1}, R_{k2}\) (and \(R^A_{k1}, R^A_{k2}\)).




Notice that the Yukawa matrices $\tilde{Y}_{1,\,2}$ are not necessarily diagonal and, as a consequence, this model allows for flavor-changing neutral currents at tree level.

\subsection{Charged Yukawa Lagrangian}
As far as the charged part of the Yukawa Lagrangian is concerned, it also is  presented in a compact expression as follows:
\begin{equation}
\begin{split}
\mathcal{L}_{\text{Yukawa}}^{\text{charged}} = \epsilon_{f}\sec\beta \bigg[ &-\frac{\sqrt{2}\tan\beta}{v} \bar{M}^{f} \delta_{ij} \left( H^{+} \bar{f}_{i} g_{i} + H^{-} \bar{g}_{i} f_{i} \right) \\
&+ \tilde{Y}_{2,ij}^{f} \left( H^{+} \bar{f}_{i} g_{j} + H^{-} \bar{g}_{j} f_{i} \right) \bigg]
\end{split}
\end{equation}
where $f$ denotes the up-type fermion and $g$ the down-type fermion in each corresponding doublet (i.e. $(f,g) = (u,d)$ for quarks, and $(\nu,\ell)$ for leptons), and $\tilde Y_{2,ij}^{f}$ are the corresponding rotated matrices defined previously.

\subsection*{Explicit expressions for the mixing matrices}

In the basis $(h_1,h_2,h_{\eta 1},h_{\eta 2})$, the CP-even mixing matrix $R$ is given by

\[
R^S = 
\begin{pmatrix}
-\sin\alpha & \cos\alpha & 0 & 0 \\
\cos\alpha & \sin\alpha & 0 & 0 \\
0 & 0 & \cos\theta_\eta & \sin\theta_\eta \\
0 & 0 & -\sin\theta_\eta & \cos\theta_\eta
\end{pmatrix},
\]

while in the basis $(a_1,a_2,a_{\eta 1},a_{\eta 2})$, the CP-odd mixing matrix $R^A$ is

\[
R^A = 
\begin{pmatrix}
-\sin\beta & \cos\beta & 0 & 0 \\
\cos\beta & \sin\beta & 0 & 0 \\
0 & 0 & \cos\theta_{\eta A} & \sin\theta_{\eta A} \\
0 & 0 & -\sin\theta_{\eta A} & \cos\theta_{\eta A}
\end{pmatrix}.
\]

The corresponding elements and the auxiliary functions $\Delta R_k^S$ and $\Delta R^A_k$ are summarized in Tables~\ref{Ss} and~\ref{As}.

\begin{table}[h]
\centering
\caption{CP-even sector: elements of $R^S$ and $\Delta R_k^S$.}~\label{Ss}
\begin{tabular}{c|cccc|c}
\hline
State & $R_{k1}^S$ & $R_{k2}^S$ & $R_{k3}^S$ & $R_{k4}^S$ & $\Delta R_k^S$ \\
\hline
$S_1$ & $-\sin\alpha$ & $\cos\alpha$ & 0 & 0 & $\dfrac{\cos(\alpha-\beta)}{\cos\beta}$ \\
$S_2$ & $\cos\alpha$ & $\sin\alpha$ & 0 & 0 & $\dfrac{\sin(\alpha-\beta)}{\cos\beta}$ \\
$S_3$ & 0 & 0 & $\cos\theta_\eta$ & $\sin\theta_\eta$ & 0 \\
$S_4$ & 0 & 0 & $-\sin\theta_\eta$ & $\cos\theta_\eta$ & 0 \\
\hline
\end{tabular}
\end{table}

\begin{table}[h]
\centering
\caption{CP-odd sector: elements of $R^A$ and $\Delta R^A_k$.}\label{As}
\begin{tabular}{c|cccc|c}
\hline
State & $R^A_{k1}$ & $R^A_{k2}$ & $R^A_{k3}$ & $R^A_{k4}$ & $\Delta R^A_k$ \\
\hline
$A_1$ & $-\sin\beta$ & $\cos\beta$ & 0 & 0 & $\dfrac{1}{\cos\beta}$ \\
$A_2$ & 0 & 0 & $\cos\theta_{\eta A}$ & $\sin\theta_{\eta A}$ & 0 \\
$A_3$ & 0 & 0 & $-\sin\theta_{\eta A}$ & $\cos\theta_{\eta A}$ & 0 \\
\hline
\end{tabular}
\end{table}


\subsection{Modified Higgs-Gauge Boson Couplings in the \(D_4\) Model}

In the \(D_4\) model, the flavon doublet \(\eta = (\eta_1,\eta_2)^T\) is an SM gauge singlet (it carries no  SU(2)\(_L \times\) U(1)\(_Y\) quantum numbers). Consequently, only the two Higgs doublets \(\Phi_1\) and \(\Phi_2\) couple directly to the \(W\) and \(Z\) bosons through their kinetic terms
\[
\mathcal{L}_{\rm kin} = |D_\mu \Phi_1|^2 + |D_\mu \Phi_2|^2.
\]

After electroweak symmetry breaking, the gauge-boson masses are generated solely by the Higgs VEVs:
\[
M_W^2 = \frac{g^2}{4}(v_1^2 + v_2^2), \qquad
M_Z^2 = \frac{M_W^2}{\cos^2\theta_W},
\]
where \(v = \sqrt{v_1^2 + v_2^2} \approx 246\,\text{GeV}\). The flavon VEVs \(v_{\eta1}\), \(v_{\eta2}\) do not contribute to \(M_W\) or \(M_Z\).

The tree-level couplings of each physical scalar \(S_k\) to the gauge bosons arise exclusively from the \(h_1\) and \(h_2\) components (the flavon components \(h_{\eta1}\) and \(h_{\eta2}\) carry no gauge charge). Expanding the kinetic terms yields the interaction Lagrangian
\begin{eqnarray}
 \mathcal{L}_{\rm int}^{VV} &\supset& \frac{g^2}{2} (v_1 h_1 + v_2 h_2) W_\mu^+ W^{-\mu}\nonumber\\
&+& \frac{g^2 + g'^2}{4} (v_1 h_1 + v_2 h_2) Z_\mu Z^\mu .   
\end{eqnarray}

Substituting the mixing gives the modified couplings
\begin{equation}\label{HiggsVV_coupling}
g_{S_k W^-W^+} = \frac{g^2 M_W}{2} \,\xi_k, \qquad
g_{S_k ZZ} = \frac{g^2 M_Z}{2 \cos^2\theta_W} \,\xi_k,    
\end{equation}
with the explicit modification factors
\[
\xi_k = \frac{v_1 R_{k1} + v_2 R_{k2}}{v} = \cos\beta \, R_{k1} + \sin\beta \, R_{k2}.
\]
In the SM limit, one recovers \(\xi_k = 1\) for the SM-like Higgs and \(\xi_k = 0\) for all other scalars. In the \(D_4\) model the \(\xi_k\) varies from 0 or 1 because of Higgs–flavon mixing encoded in the orthogonal matrix \(R\).

For the CP-odd and charged scalars, the couplings to gauge bosons follow the standard 2HDM pattern (controlled by \(\tan\beta\)) and receive no additional tree-level modification from the flavon.

\section{Model parameter space}\label{sec:SecIII}
Once the model has been developed, we now proceed to analyze a subset of the full parameter space of the model. We are interested in exploring the predictive power of the model, in particular FCNC via the decays $h\to \tau^-\mu^++\tau^+\mu^-$ (shortened to $h\to \tau\mu$). The model parameters that have a direct impact on these processes are the following: 
\begin{itemize}
    \item Cosine of the difference of the mixing angles $\alpha$ and $\beta$: $\cos(\alpha-\beta)$,
    \item Ratio of the VEVs of the scalar doublets $v_2/v_1=\tan\beta$,
    \item The rotated Yukawa matrix element $[\tilde{Y}_2]_{32}^{\ell}=[\tilde{Y}_2]_{23}^{\ell}$,
    \item The cutoff scale $\Lambda$,
    \item The VEV $v_{\eta1,2}$,
    \item Parameters of the scalar potential (Eq.~\eqref{ScalarPotential}) and coming from the rotation matrix $\mathcal{O}_f$ (Eq.~\eqref{MatrixO}).
\end{itemize}
We bound these parameters through theoretical and experimental constraints as shown in Table~\ref{observables}, where each constraint considered is presented in the following subsections \eqref{TC}-\eqref{ExpC}.
\begin{table}[!htb]

\caption{Free model parameters and their constraints, including LHC Higgs signal strengths $\mu_X$, excess of events in resonances (EER), and upper limits (U.L.) lepton flavor violation processes (LFVp).}\label{observables}

\begin{centering}
\begin{tabular}{|c|c|}
\hline 
Parameter & Constraint\tabularnewline
\hline 
\hline 
$\cos(\alpha-\beta)$ & $\mu_X$ and LFVp\tabularnewline
\hline 
$\tan\beta$ & $\mu_X$, EER and LFVp\tabularnewline
\hline 
$\ensuremath{[\tilde{Y}_{2}]_{32}^{\ell}=[\tilde{Y}_{2}]_{23}^{\ell}}$ & U.L. on $\mathcal{BR}(h\to\tau\mu)$ and $\mathcal{BR}(\tau\to\mu\gamma)$\tabularnewline
\hline 
Cutoff scale $\Lambda$ & $\mu_X$ and LFVp\tabularnewline
\hline 
VEVs $v_{\eta1},v_{\eta2}$ & $\mu_X$, EER and LFVp \tabularnewline
\hline 
\end{tabular}
\par\end{centering}
\end{table}

\begin{figure}[!htb]
		\centering
		\includegraphics[width=\linewidth]{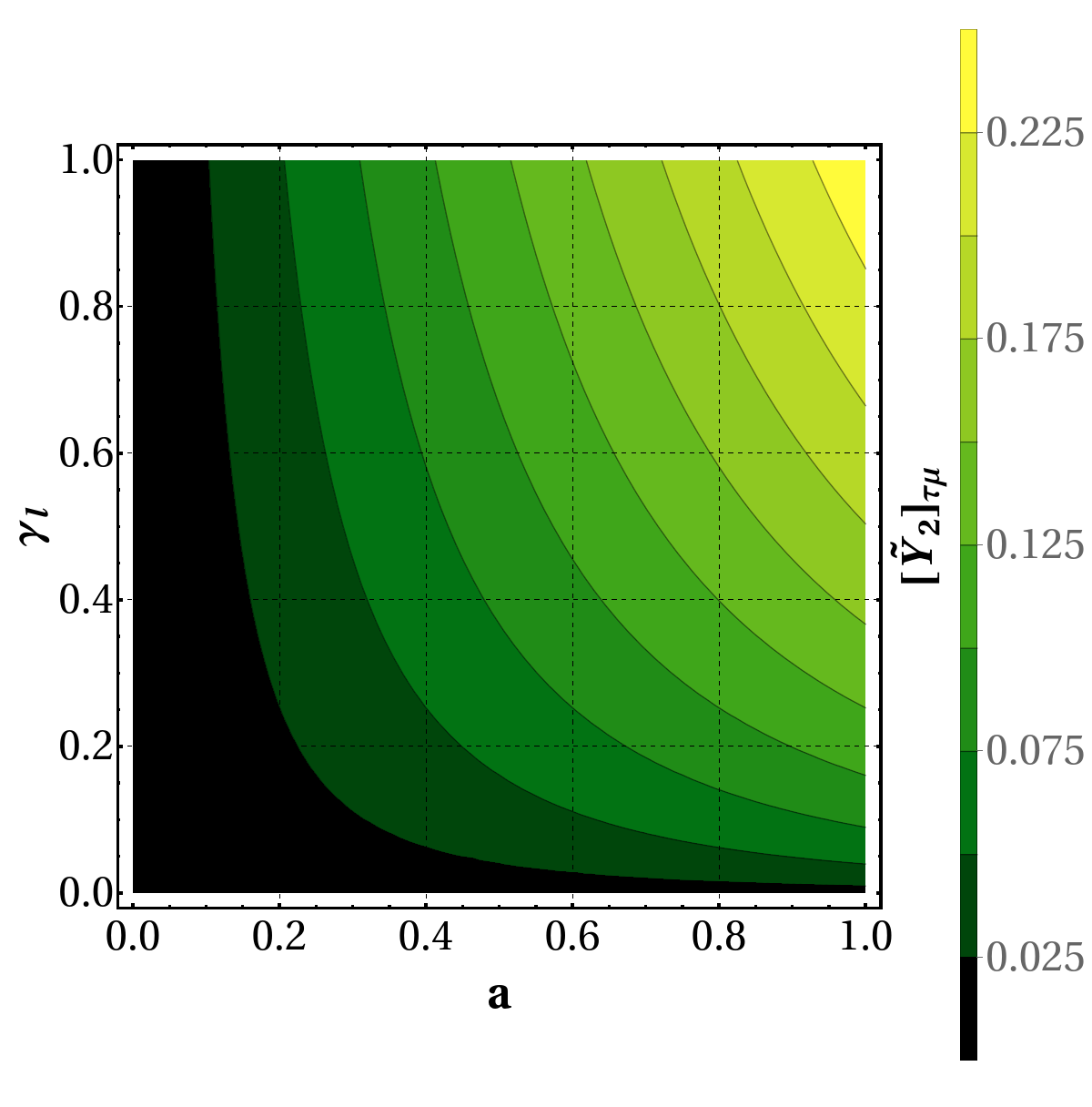}
		\caption{Values of $[\tilde{Y}_2]_{\tau\mu}$ as a function of $\gamma_\ell$ and $a$. }\label{gamma-a}
	\end{figure}

According to the Yukawa Lagrangian in Eq.~\eqref{FullYuwLag}, the evaluation of the flavor-violating observables requires the explicit form of the rotated Yukawa matrix \([\tilde{Y}_2^f]_{ij}\). In what follows, we focus on Case 15 to scrutinize the specific matrix elements that govern the \(g_{h\tau \mu }\) coupling, given that the leptonic decay \(h\to\tau\mu\) constitutes the benchmark prediction of this work. This particular scenario offers the most minimal and predictive expression: the primary Yukawa matrix \(Y_{2}\) is strictly sparse, with its sole non-vanishing entry located in the third-generation diagonal component, namely \([Y_2]_{33}=a\). Upon imposing the transformation criteria outlined in Eq.~\eqref{rotated_Yukawas}, the off-diagonal elements in the rotated basis emerge as
\begin{equation}\label{Y2_23-rotada}
[\tilde{Y}_2]_{23}^f = [\tilde{Y}_2]_{32}^f = \frac{a\, m_2}{m_3-m_2} \sqrt{\gamma_f(r_{32}-\gamma_f-1)},
\end{equation}
where the massless limit for the first generation (\(m_1\to 0\)) has been invoked, and the mass ratio is defined as \(r_{32} \equiv m_3/m_2\). Finally, the parametric dependence of the specific flavor-changing coupling \([\tilde{Y}_2]_{\tau\mu}\) as a function of the scaling factor \(\gamma _{\ell }\) and the coupling strength \(a\) is illustrated in Fig.~\ref{gamma-a}.

\subsection{Experimental constraints}\label{ExpC}
\subsubsection{Signal strengths}
 We first confront the free parameters of the model against LHC Higgs signal strengths ($\mu_X$), defined as follows:
 \begin{equation}\label{muX}
		\mathcal{\mu}_{X}=\frac{\sigma(pp\to h)\cdot\mathcal{BR}(h\to X)}{\sigma(pp\to h^{\text{SM}})\cdot\mathcal{BR}(h^{\text{SM}}\to X)},
	\end{equation}
	where $\sigma(pp\to S)$ is the production cross-section of $S$, with $S=h,\,h^{\text{SM}}$; here $h$ is the SM-like Higgs boson coming from an extension of the SM and $h^{\text{SM}}$ is the SM Higgs boson; $\mathcal{BR}(S\to X)$ is the branching ratio of the decay $S\to X$, with $X=c\bar{c},\,b\bar{b},\;\tau^-\tau^+,\;\mu^-\mu^+,\;WW^*,\;ZZ^*,\;\gamma\gamma$. 
    \\
	 From the Yukawa Lagrangian in Eq.~\eqref{FullYuwLag}, the coupling of the SM-like Higgs boson $S_1\equiv h$ to fermions $f_i$ and $f_j$ is extracted as

\begin{equation}\label{yukawacoupling}
  g_{h f_i f_j} = 
-\frac{m_f}{v}\frac{\sin\alpha}{\cos\beta}\delta_{ij}
+ \frac{\cos(\alpha-\beta)}{\sqrt{2}\cos\beta} \, \Bigg(\tilde{Y}_{2,ij}^{f}
+ \mathcal{F}^f_{ij}
\Bigg).
\end{equation}
The first term is flavor-diagonal and gives the SM coupling in the alignment limit $\alpha = \beta - \pi/2$, where $\sin\alpha = -\cos\beta$. The second term is non-diagonal flavor and is proportional to $\cos(\alpha-\beta)$, disappearing when the alignment limit is exact. The second term is responsible for FCNC at the tree level via the non-diagonal rotated Yukawa matrix $\tilde{Y}_{2,ij}^f$. This contribution induces BSM processes such as Higgs boson flavor-violating processes  (e.g. $h \to f_if_j$), with a strength dependent on the parameters of the four-zero Yukawa textures. The interactions $hW^-W^+$ and $hZZ$ are given in Eq.~\eqref{HiggsVV_coupling}. The $X=\gamma\gamma$ channel needs special treatment since it arises at the one-loop level, with contributions from fermions, $W^\pm$ gauge bosons and charged scalar bosons $H^\pm$ --- the latter arising from the second scalar doublet $\Phi_2$ ---. Analytical expressions for the decay width $\Gamma(h\to\gamma\gamma)$ are given in Eq.~\eqref{h-gam_gam}.

\begin{equation}\label{h-gam_gam}
\Gamma(h\to\gamma\gamma)=\frac{\alpha^2 M_h^3}{1024\pi^3 M_W^2}\Bigg|\sum_{s}A_s^{h\gamma\gamma}(\tau_s)\Bigg|^2,
\end{equation}
where the subscript $s=0,1/2,1$ refers to the spin of the charged particle circulating in the loop, and 
\begin{align*}
A_{s}^{h\gamma\gamma} = 
\begin{cases}
    \displaystyle \sum_{f}\frac{2m_{W}g_{h}^{f}N_{c}Q_{f}^{2}}{m_{f}}\Big[-2\tau_{f}\big(1+(1-\tau_{f})f(\tau_{f})\big)\Big] \\[1ex] 
    \hspace{4cm} \text{for } s=\frac{1}{2}, \\[2ex]
    \displaystyle g_{hWW}\Big[2+3\tau_{W}+3\tau_{W}(2-\tau_{W})f(\tau_{W})\Big] \\[1ex] 
    \hspace{4cm} \text{for } s=1, \\[2ex]
    \displaystyle \frac{m_{W}g_{hH^{-}H^{+}}}{M_{H^{\pm}}^{2}}\Big[\tau_{H^{\pm}}(1-\tau_{H^{\pm}}f(\tau_{H^{\pm}}))\Big], \\[1ex] 
    \hspace{4cm} \text{for } s=0,
\end{cases}
\end{align*}
where $N_{c}=3 (1)$ for quarks (leptons), $\tau_a=4M_a^2/M_h^2$, and 
\begin{align*}
	f(x)=
 \begin{cases}
        \displaystyle{\arcsin^2\Bigg(\frac{1}{\sqrt{x}}\Bigg),}& x\geq1,\\
		\displaystyle{-\frac{1}{4}\Bigg[\log\Bigg(\frac{1+\sqrt{1-x}}{1-\sqrt{1-x}}\Bigg)-i\pi\Bigg]^{2}},&x<1.
  \end{cases}
  \end{align*}

\subsubsection{Radiative decays $\ell_i\to\ell_j\gamma$}
\begin{figure}[!htb]
		\centering
		\includegraphics[width=6cm]{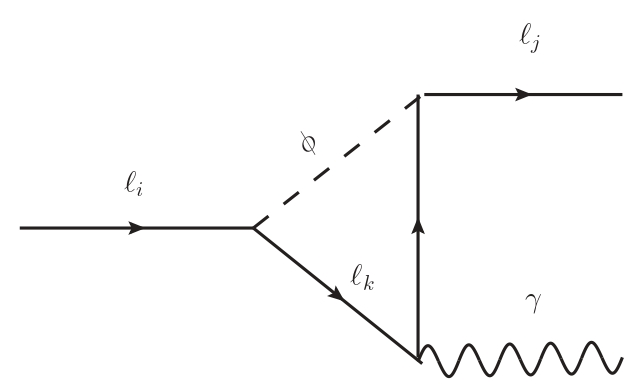}
		\caption{Feynman diagram of the process $\ell_i\to\ell_j\gamma$ at one-loop level induced by $\phi=S_k,\,A_k$.}\label{FDliljgamma}
	\end{figure}

The effective Lagrangian for the $\ell_i\to\ell_j\gamma$ process is given by
	\begin{equation}\label{EffLagrangian}
		\mathcal{L}_{\text{eff}}=C_L Q_{L\gamma} C_R Q_{R\gamma}+h.c.,
	\end{equation}
	where the dim-5 electromagnetic penguin operators read
	\begin{equation}\label{dim5Op}
		Q_{L\gamma,\,R\gamma}=\frac{e}{8\pi^2}(\bar{\ell}_j\sigma^{\alpha\beta}P_{L,\,R}\ell_i)F_{\alpha\beta},
	\end{equation}
	where $F_{\alpha\beta}$ is the electromagnetic field strength tensor. The Feynman diagram of the process $\ell_i\to\ell_j\gamma$ is depicted in Fig. \ref{FDliljgamma}.

	The Wilson coefficients $C_{L,\,R}$ receive contributions at one-loop level and an important contribution from the Barr-Zee two-loops level. For the particular case where $\ell_i=\tau$ and $\ell_j=\mu$, the approximations $g_{\phi\mu\mu}\ll g_{\phi\tau\tau}$ and $m_\mu\ll m_{\tau}\ll m_\phi$ are assumed. According to this assertion, the one-loop Wilson coefficients $C_{L,\,R}$ simplify to the expressions shown in Eq.~\eqref{WilsonCoe1loop}~\footnote{For completeness, we include contributions from $S_{3,4}$ and $A_{2,3}$. However, these flavon particles do not contribute as $g_{S_{3,4}\tau\mu}$ and $g_{A_{2,3}\tau\mu}$ vanish, as shown in Tables~\ref{Ss}-\ref{As}.} \cite{Harnik:2012pb, Blankenburg:2012ex}.
	
    \begin{align}\label{WilsonCoe1loop}
    C_L^{1-\text{loop}} &\simeq \sum_{\phi=S_k,\,A_k}\frac{g_{\phi\tau\tau}^{*}g_{\phi\tau\mu}^{*}}{12m_\phi^2}\Bigg(-4+3\log\frac{m_\phi^2}{m_{\tau}^2}\Bigg),\notag \\
    C_R^{1-\text{loop}} &\simeq\sum_{\phi=S_k,\,A_k}\frac{g_{\phi\tau\tau}g_{\phi\mu\tau}}{12m_\phi^2}\Bigg(-4+3\log\frac{m_\phi^2}{m_{\tau}^2}\Bigg).
    \end{align}

	The numerical expressions for 2-loop contributions are given by
	
	\begin{align}\label{WilsonCoe2loop}
		C_L^{2-\text{loops}}&=\sum_{\phi=S_k,\,A_k}g_{\phi\tau\mu}^*(-0.082 g_{\phi tt}+0.11)/(m_\phi)^2, \nonumber \\
		C_R^{2-\text{loops}}&=C_L^{2-loops}(g_{\phi\tau\mu}^*\to g_{\phi\tau\mu}).
	\end{align}
	
	The rate for $\tau\to\mu\gamma$ is
	
	\begin{equation}\label{ratetaumugamma}
		\Gamma(\tau\to\mu\gamma)=\frac{\alpha m_{\tau}^2}{64\pi^4}(|C_L|^2+|C_R|^2).
	\end{equation}
    

The corresponding decay width of the process $\mu\to e\gamma$ is obtained by replacing $\tau\to \mu,\, \mu\to e$ from \eqref{EffLagrangian} to \eqref{ratetaumugamma}, while for the process $\tau\to e\gamma$, replacing $\mu\to e$ is required.


\subsubsection{$h\to \ell_i \ell_j$ decays}\label{hlilj}
	
\begin{figure}[!htb]
		\centering
		\includegraphics[width=5cm]{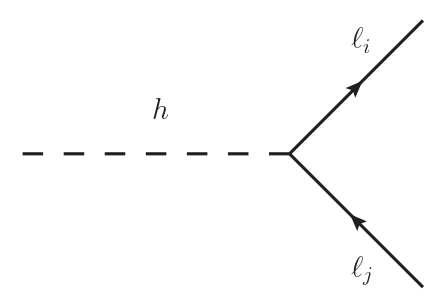}
		\caption{Feynman diagram of the process $h\to\ell_i\ell_j$ at tree level.}\label{FDhlilj}
	\end{figure}

	The LFV processes $h\to \ell_i\ell_j$ ($\ell_{i,\,j}=\ell_{i,\,j}^-\ell_{i,\,j}^+$) where $\ell_i\ell_j=e\mu,\,e\tau,\,\tau\mu$ can arise at tree level in many models that extend to the SM, as shown in Fig. \ref{FDhlilj}.
	
	The relevant interactions can be extracted from the Yukawa Lagrangian
	\begin{equation}
		\mathcal{L}_Y\supset -Y_{ij}\bar{\ell}_L^i \ell_R^jh+h.c.
	\end{equation}	 
	The corresponding full decay width of the $h\to \bar{f}_i f_j$ decays is given by:
	\begin{align}\label{Wid_h-lilj}
		\Gamma(h\to \bar{f}_i f_j)=\frac{N_c g^2_{h \bar{f}_i f_j }m_h}{128\pi}\Bigg[ 4-(\sqrt{\tau_{f_i}} \notag \\
        +\sqrt{\tau_{f_j}})^2  \Bigg]^{3/2}\sqrt{4-(\sqrt{\tau_{f_i}}-\sqrt{\tau_{f_j}})^2},
	\end{align}
	where $g_{h \bar{f}_i f_j}$ is the $h\bar{f}_i f_j$ coupling coming from an extension of the SM, $N_c=3\,(1)$ is the color number for quarks (leptons), $m_h$ is the Higgs boson mass and $\tau_{f_i}=4m_{f_i}^2/m_h^2$. 

    	\subsubsection{$\ell_i\to\ell_j\ell_k\bar{\ell}_k$ decays}
\begin{figure}[!htb]
		\centering
		\includegraphics[width=5cm]{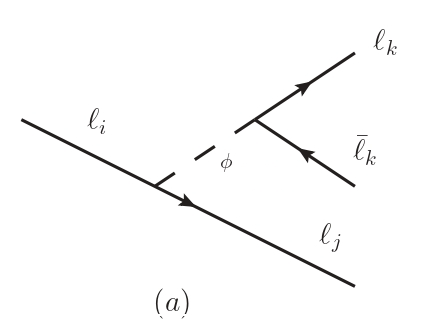}
		\includegraphics[width=5cm]{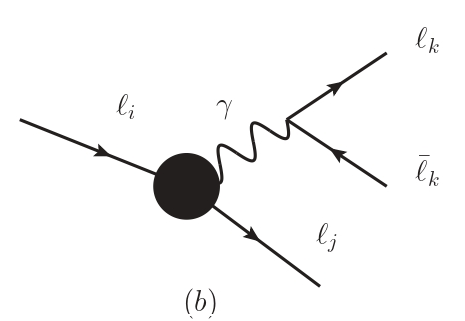}
		\caption{Feynman diagrams that contribute to the $\ell_i\to \ell_j\ell_k\bar{\ell}_k$, (a) tree-level and (b) one-loop level, where the black circle represents the loop as Feynman diagram of Fig. \ref{FDliljgamma}}\label{FDliljlklk}
	\end{figure}

	Within the 2HDM-Tx framework, these type of decays can be induced at tree-level through the exchange of $S_k$ and $A_k$. The Feynman diagrams are depicted in Fig. \ref{FDliljlklk}. However, this process is suppressed by the flavor-violating Yukawa couplings $\tilde{Y}_{\ell_i\ell_j}$ and by the flavor-conserving coupling $\tilde{Y}_{\ell_k\ell_k}$, where $\ell_k=e,\,\mu$. Nevertheless, there are higher order contributions at one-loop level, slightly complementing the constraints --- more restrictive --- analyzed above. The corresponding flavor-violating partial decay width reads
	\begin{equation}\label{Wid_tau-3mu}
		\Gamma(\tau\to3\mu)\sim \frac{\alpha m_{\tau}^5}{6(2\pi)^5}\Big|\log\frac{m_{\mu}^2}{m_{\tau}^2}-\frac{11}{4} \Big|(|C_L|^2+|C_R|^2),
	\end{equation}  
where the suppressed terms by the muon mass are neglected. The Wilson coefficients $C_{L,\,R}$ are given in Eqs. \eqref{WilsonCoe1loop}-\eqref{WilsonCoe2loop}. 

\subsection{Theoretical constraints}\label{TC}
From the scalar potential in Eq.~\eqref{ScalarPotential}, we derive constraints on the parameters that build the potential.  
In order to maintain the stability of the scalar potential~\eqref{ScalarPotential}, all the following constraints must be satisfied simultaneously:
\begin{eqnarray}
&&    \lambda_1 > 0, \quad \lambda_2 > 0, \quad \lambda_\eta > 0,\quad\lambda_3 + \lambda_4 > -2\sqrt{\lambda_1\lambda_2},\nonumber\\
&&\qquad\quad    \lambda_{\Phi\eta} > -2\sqrt{\lambda_1 \lambda_\eta}, \quad
\lambda_{\Phi\eta} > -2\sqrt{\lambda_2 \lambda_\eta}.
\end{eqnarray}

The tree-level unitarity restrictions are as follows.
\begin{eqnarray}
  &&  \lambda_1 < 8\pi, \quad \lambda_2 < 8\pi, \quad \lambda_\eta < 8\pi,\nonumber\\
&&\quad|\lambda_3 + \lambda_4| < 8\pi,\quad
|\lambda_{\Phi\eta}| < 4\pi.
\end{eqnarray}

The cubic terms \(\kappa_1\) and \(\kappa_2\) terms are irrelevant for stability and vanish in the high-energy limit and do not enter the unitarity bounds.

\begin{figure*}[!htb]
    \centering
    \subfigure[]{\includegraphics[width=0.47\linewidth]{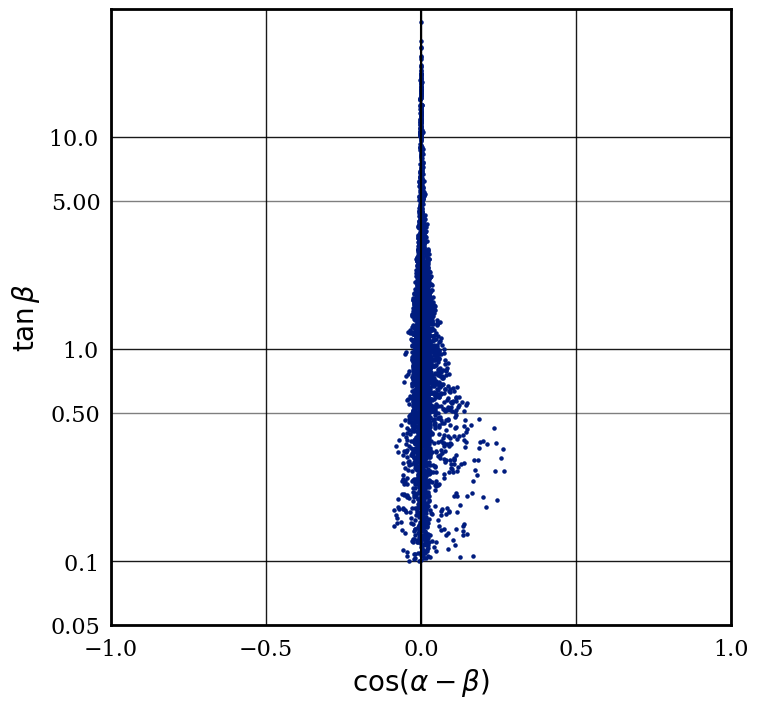}}
    \hfill
    \subfigure[]{\includegraphics[width=0.47\linewidth]{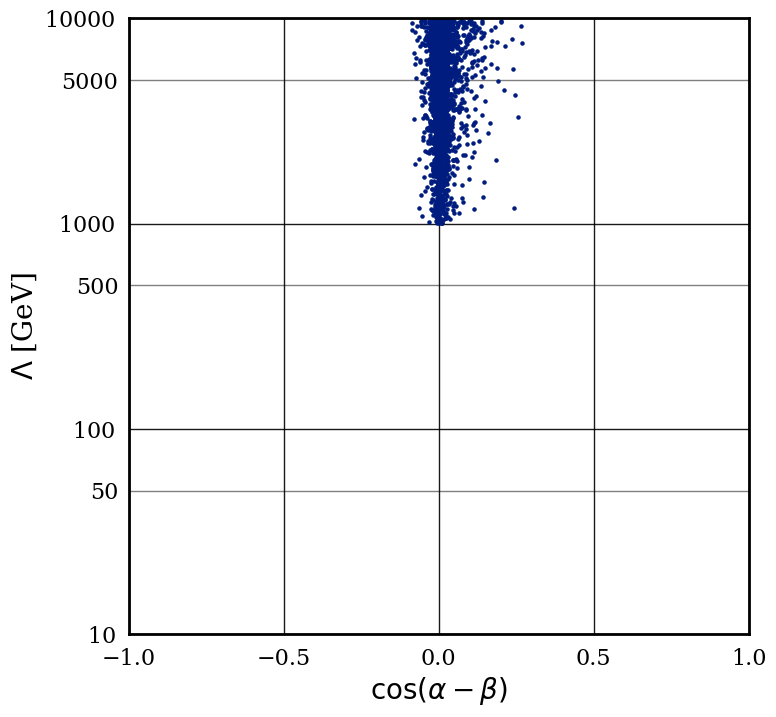}}
    \\
    \subfigure[]{\includegraphics[width=0.47\linewidth]{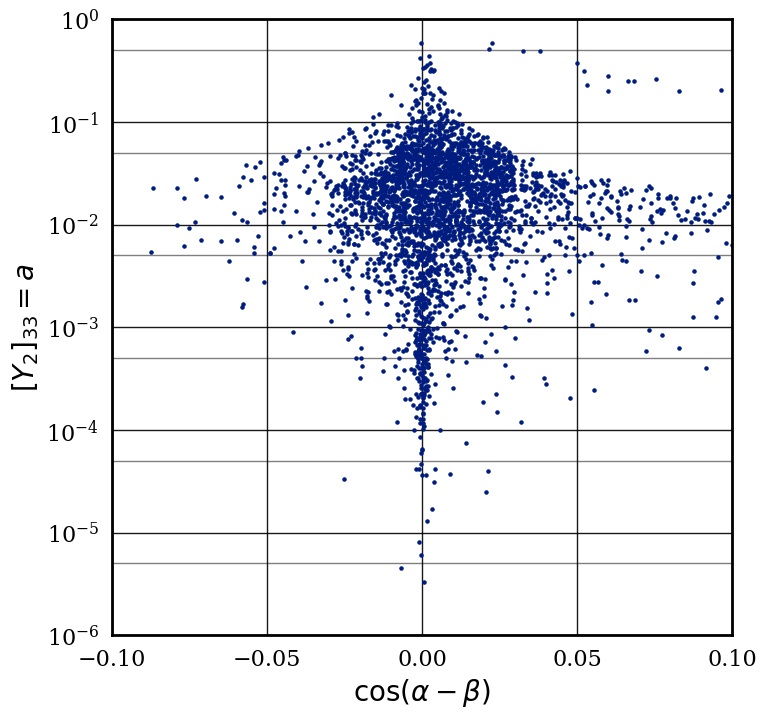}}
    \hfill
    \subfigure[]{\includegraphics[width=0.47\linewidth]{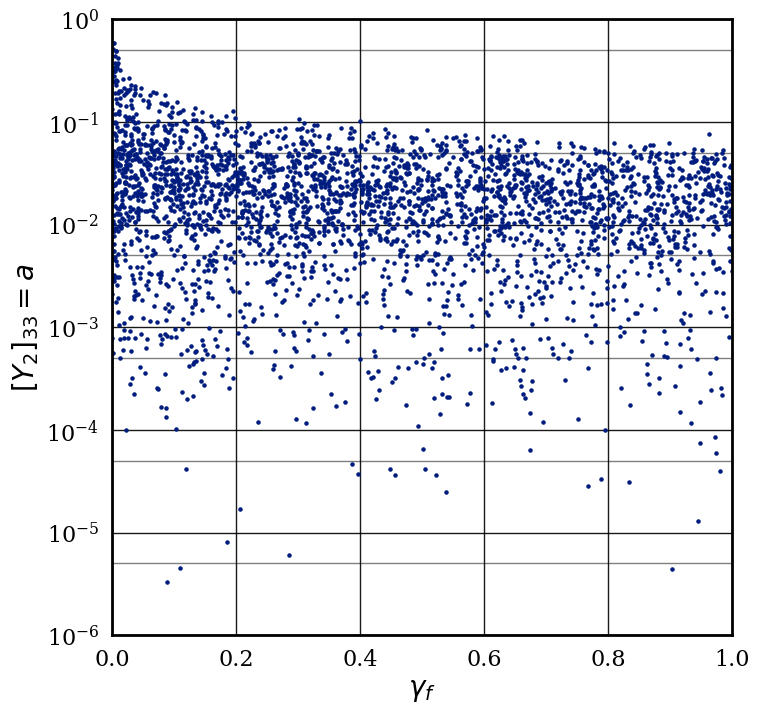}}
    \caption{Allowed parameter space: $a)$ $\cos(\alpha-\beta)-\tan\beta$ plane, $b)\,\cos(\alpha-\beta)-\Lambda$ plane, $c)$ $\cos(\alpha-\beta)$ versus Yukawa matrix element $[Y_2]_{33}=a$ plane, and $d)$ $\gamma_f$-$a$ plane. The blue points are those that satisfy all the constraints considered in this section.}
    \label{Cab-tb_plane}
\end{figure*}

Having presented the observables that enable us to constrain the free parameters of the model, we proceed with a numerical analysis through a systematic scan over the parameter ranges listed in Table~\ref{scan}. For this purpose, we focus on a specific benchmark scenario, namely case 15 defined in Eq.~\eqref{caso15}. This analysis yields stringent constraints on several of the model's free parameters. Figure~\ref{Cab-tb_plane} displays four projection planes: $a)$ $\cos(\alpha-\beta)$ versus $\tan\beta$; $b)$ $\cos(\alpha-\beta)$ versus $\Lambda$; $c)$ $\cos(\alpha-\beta)$ as a function of the Yukawa matrix element $[Y_2]_{33}=a$; and $d)$ $\gamma_f$ versus $a$. In particular, the $\cos(\alpha-\beta)$-$\tan\beta$ plane reveals a preferred region characterized by low values of $\cos(\alpha-\beta)\sim 0$ for $5\lesssim\tan\beta\lesssim 40$, despite having explored approximately $5\times 10^{10}$ random points that successfully satisfy all the aforementioned theoretical and experimental constraints. This finding is particularly significant, given that the 2HDMs typically impose no upper bound on $\tan\beta$; the observed bound here originates from the excess of events observed in the resonances at 95 and 152~GeV. For $\tan\beta\lesssim 5$, the allowed range of $\cos(\alpha-\beta)$ broadens considerably, extending up to 0.25. This behavior is noteworthy, as $\cos(\alpha-\beta)$ constitutes the primary source of FCNCs and simultaneously governs the rate of the process of interest, namely $h\to\tau\mu$. Additional parameters that turn out to be heavily constrained are illustrated in the $\gamma_f$-$a$ plane. Here, $a$ denotes the $(3,3)$ entry of the Yukawa matrix $Y_2$ for case 15, while $\gamma_f$ parametrizes the $(3,3)$ element---denoted $A_f$---of the mass matrix defined in Eqs.~\eqref{MassMatrixElementA}.

\begin{figure}
    \centering
    \includegraphics[width=\linewidth]{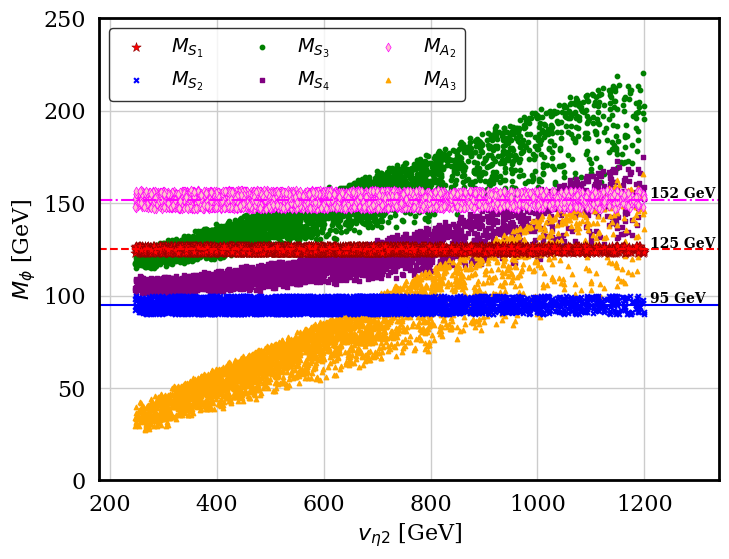}
    \caption{CP-even masses ($M_{S_1}=M_h, M_{S_2}, M_{S_3}, M_{S_4}$) and CP-odd masses ($M_{A_2}, M_{A_3}$) against the values of the vacuum-expectation-value $v_{\eta 2}$. We identify $S_1$ as the SM Higgs, while $M_{S_2}$($M_{A_2}$) are associated with the 95 (152) GeV excesses.}
    \label{fig:masses-vs-veta2}
\end{figure}

The values of the CP-even scalar masses $M_{S_1}$, $M_{S_2}$, 
$M_{S_3}$, and $M_{S_4}$, together with the CP-odd masses 
$M_{A_2}$ and $M_{A_3}$, for the allowed parameter points are 
shown as a function of the flavon vacuum expectation value 
$v_{\eta_2}$ in Fig.~\ref{fig:masses-vs-veta2}. We identify $S_1$ 
with the SM-like Higgs boson and require $M_{S_1}\simeq 125$~GeV, 
while $S_2$ and $A_2$ are associated with the reported excesses 
around $95$~GeV and $152$~GeV, respectively. The numerical scan 
shows that these three mass requirements can be simultaneously achieved for a broad range of $v_{\eta_2} \in [v,1200] \text{ GeV}$. Meanwhile, the 
remaining masses $S_3$, $S_4$, and $A_3$, which are 
not fixed by these phenomenological requirements, exhibit a clear 
proportional dependence on the flavon VEV, in particular $v_{\eta_2}$, although a 
sizable spread remains as a consequence of the simultaneous 
variation of the scalar-potential parameters. This is due to the explicit dependence of the CP-even and CP-odd mass matrices on $v_{\eta_1}, v_{\eta_2}$, indicating 
that the flavon symmetry-breaking scale plays an important role 
in determining the masses of the additional scalar states. The values of these masses for the parameter scan performed are $M_{S_3} \in [114,220] \text{ GeV}$, $M_{S_4} \in [100,174] \text{ GeV}$ and $M_{A_3} \in [27,166] \text{ GeV}$.

\begin{figure}
    \centering
    \includegraphics[width=\linewidth]{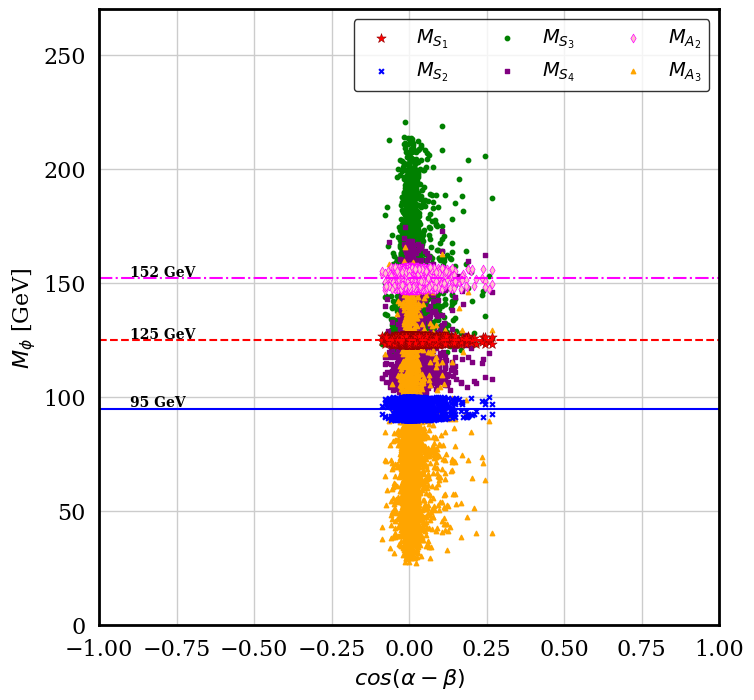}
    \caption{CP-even masses ($M_{S_1}(\equiv M_h), M_{S_2}, M_{S_3}, M_{S_4}$) and CP-odd masses ($M_{A_2}, M_{A_3}$) as a function of $\cos (\alpha-\beta)$. $M_{S_2}$($M_{A_2}$) are associated with the 95 (152) GeV excesses.}
    \label{fig:masses-cosab}
\end{figure}

The dependence of the scalar spectrum on the mixing-angle combination
$\cos(\alpha-\beta)$ is shown in Fig.~\ref{fig:masses-cosab}. In particular, the limit $\cos(\alpha-\beta)\to 0$ corresponds to the suppression of the FCNC contribution, as seen in Eq. \eqref{yukawacoupling}. The scanned points are mainly found in the region close to
$\cos(\alpha-\beta)=0$, approximately within
$-0.1\lesssim\cos(\alpha-\beta)\lesssim0.3$. Nevertheless, the
requirements $M_{S_1}\simeq125$~GeV, $M_{S_2}\simeq95$~GeV, and
$M_{A_2}\simeq152$~GeV can also be simultaneously satisfied for
non-vanishing values of $\cos(\alpha-\beta)$, thus not requiring
the flavor-conserving limit, leaving a region of parameter space in
which potentially relevant flavor-changing neutral scalar interactions
can be present.

\begin{table}
\caption{Parameters and their scanned range.}\label{scan}
\begin{centering}
\begin{tabular}{|c|c|}
\hline 
Parameter & Scanned range\tabularnewline
\hline 
\hline 
$\lambda_{1,2,3,4,\eta,\phi\eta},\,\kappa_{1,2}$ & $[0,\,0.01]$\tabularnewline
\hline 
$v_{\eta1,\eta2}$ & $[v,\,1200]$ GeV\tabularnewline
\hline 
$\tan\beta$ & $[0.1,\,50]$\tabularnewline
\hline 
$\cos(\alpha-\beta)$ & $[-1,\,1]$\tabularnewline
\hline 
$\Lambda$ & $[1000,\,10000]$ GeV\tabularnewline
\hline 
$a,\,\gamma$ & $[0,\,1]$\tabularnewline
\hline 
$M_{H^{\pm}}$ & $[110,\,140]$ GeV\tabularnewline
\hline 
$M_{A_{1}}$ & $[300,\,1000]$ GeV\tabularnewline
\hline 
\end{tabular}
\par\end{centering}
\end{table}

\begin{figure}[!htb]
		\centering
		\includegraphics[width=\linewidth]{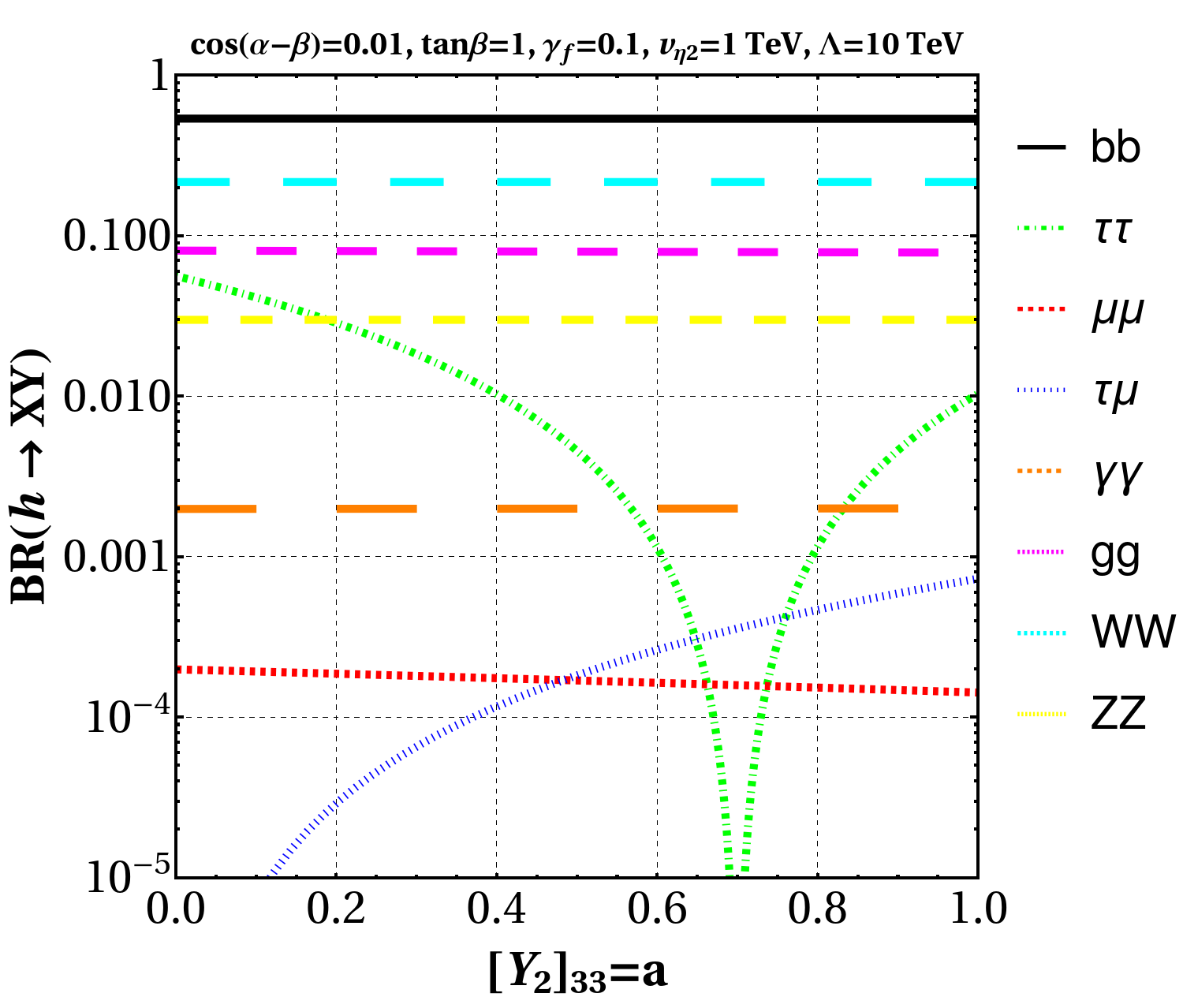}
		\caption{Branching ratio of the Higgs boson into dominant channels as a function of the matrix element $[Y_2]_{33}=a$. }\label{HiggsBRs}
	\end{figure}

\subsection{Validation of the Simulation Framework against CMS Data}

Before proceeding with the predictive collider analysis for the HL-LH at $\sqrt{s} = 14$ TeV, it is crucial to validate our computational pipeline against real experimental data. 
For this purpose, we performed a dedicated Monte Carlo simulation of the LFV Higgs boson decays, $h \to \mu\tau\to\mu e\nu_e\nu_\tau$ ($h \to \mu\tau_e$) and $h \to \mu\tau\to\mu \mu\nu_\mu\nu_\tau$ ($h \to \mu\tau_\mu$), specifically configured to match the Run 2 LHC conditions. 
The center-of-mass energy was set to $\sqrt{s} = 13$ TeV with an integrated luminosity of $137\text{ fb}^{-1}$, in strict alignment with the data-taking period reported by the CMS Collaboration~\cite{CMS-HIG-20-009}.

\begin{figure*}[htbp]
    \centering
    \includegraphics[width=0.95\textwidth]{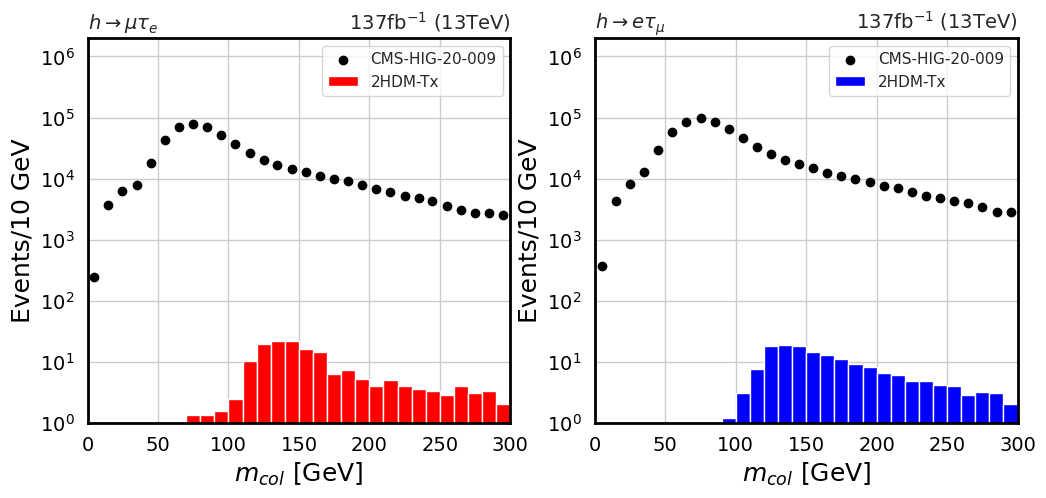} 
    \caption{Collinear mass $m_{\text{col}}$ distributions for the $h \to \mu\tau_e$ (left) and $h \to e\tau_\mu$ (right) channels. The 2HDM-Tx simulated signal (filled red/blue histograms) is superimposed on the background and data points reported by the CMS Collaboration at $\sqrt{s} = 13$ TeV with an integrated luminosity of $137\text{ fb}^{-1}$~\cite{CMS-HIG-20-009}, demonstrating the kinematic consistency of our framework.}
    \label{fig:collinear_mass_validation}
\end{figure*}

In Fig.~\ref{fig:collinear_mass_validation}, we present the collinear mass ($m_{\text{col}}$) distributions predicted by the 2HDM-Tx alongside the data and background components provided in the CMS open-access material. 
The collinear mass $m_{\text{col}}=m_{\text{vis}}/\sqrt{x_{\text{vis}}}$, where $x_{\text{vis}}$ denotes the fraction of the $p_T$ of the $\tau$ lepton that is carried by the visible products of its decay. It is computed under the collinear approximation assuming that the $\tau$-lepton decay products are Lorentz-boosted in the parent direction \cite{Ellis:1987xu}, and serves as a robust estimator of the Higgs boson mass ($M_h$). 
As illustrated, the kinematic behavior and signal injection shapes from our 2HDM-Tx framework show an excellent agreement with the experimental distributions. 
This successful cross-check verifies the reliability of our simulation cards and multivariate analysis pipeline, establishing a validated baseline that legitimizes the high-energy projections discussed in the subsequent section.

\section{Collider analysis}\label{sec:SecIV}
In this section, we compute a Monte Carlo simulation and explore the prospects for detecting the proposed signal as well as the SM background processes that obscure it. 
\subsection{Signal and SM background processes}

\begin{itemize}
\item \textbf{SIGNAL:} The target experimental signature consists in Higgs boson production through gluon fusion in proton-proton collisions. In the final state there are exactly two isolated, opposite-sign light leptons ($\mu e$ or $\mu\mu$) accompanied by missing transverse energy ($\text{MET}$) arising from the unobserved neutrinos in the $\tau$-lepton decay chain, namely $gg\to h\to\tau\mu\to \ell\mu+$MET ($\ell=e,\,\mu$). 
\item \textbf{BACKGROUND:} The dominant SM background sources capable of mimicking the signal are categorized into three main groups and simulated at the HL-LHC design energy of $\sqrt{s} = 14$ TeV:

\begin{enumerate}
    \item \textbf{Leptonic Drell-Yan Resonance ($pp \to Z/\gamma^* \to \tau^+\tau^-$):} 
    This constitutes the single largest source of background due to its massive production cross-section and its final state extremely similar to the signal. Although the majority of $\tau$-lepton pairs decay hadronically, resulting in low lepton multiplicities (0 or 1 light leptons), approximately $12.2\%$ of the events undergo mutually leptonic decays ($\tau^+\tau^- \to \ell^+\ell^-\nu\bar{\nu}\nu\bar{\nu}$). This creates a genuine dilepton plus $\text{MET}$ state final that directly populates the search region.
    
    \item \textbf{Top-Quark Production ($t\bar{t}$ and $tW$):}
    \begin{itemize}
        \item \textbf{Pair Production ($pp \to t\bar{t}$):} When both top quarks decay leptonically via $t \to b W^+ \to b \ell^+ \nu_\ell$ and $\bar{t} \to \bar{b} W^- \to \bar{b} \ell^- \bar{\nu}_\ell$, the final state yields two high-$p_T$ isolated leptons and considerable $\text{MET}$. This background is highly energetic and can significantly contaminate the high-mass tail of the transverse mass distribution.
        \item \textbf{Single Top Associated Production ($pp \to tW$):} Similarly to $t\bar{t}$, the associated production of a single top quark and a $W$ boson enters the selection stream when both the $W$ from the top decay and the companion $W$ boson decay leptonically ($tW^\mp \to b W^\pm W^\mp \to b \ell^\pm \ell^\mp \nu \bar{\nu}$). 
    \end{itemize}
    
    \item \textbf{Electroweak Diboson Production ($WW$, $ZZ$, and $WZ$):}
    \begin{itemize}
        \item \textbf{$WW$ Production ($pp \to W^+W^-$):} The fully leptonic decay of gauge boson pairs ($W^+W^- \to \ell^+\ell^-\nu\bar{\nu}$) represents an irreducible background that structurally matches the signal topology, carrying two opposite-sign leptons and intrinsic $\text{MET}$.
        \item \textbf{$ZZ$ Production ($pp \to ZZ$):} This process enters the dilepton channel when one $Z$ boson decays charged-leptonically ($Z \to \ell^+\ell^-$) and the other decays invisibly into neutrinos ($Z \to \nu\bar{\nu}$), emulating the $\text{MET}$ signature.
        \item \textbf{$WZ$ Production ($pp \to W^\pm Z$):} Although primarily a trilepton source ($W \to \ell\nu, Z \to \ell^+\ell^-$), it contaminates the signal region if one of the three leptons fails the reconstruction efficiency cuts or falls outside the geometric acceptance of the detector ($\ |\eta| > 2.5\ $), leaving exactly two identified leptons.
    \end{itemize}
\end{enumerate}

\end{itemize}

Regarding the computational framework adopted in this work, the implementation of the relevant interactions and the automated generation of the corresponding \texttt{UFO}~\cite{Degrande:2011ua} modules are carried out using $\texttt{LanHEP}$~\cite{lanhep}. Cross-section evaluations are subsequently performed with $\texttt{MadGraph5\_aMC@NLO}$~\cite{Alwall:2014hca} at parton level. Within this simulation pipeline, the parton-shower evolution and subsequent hadronization processes are handled by $\texttt{Pythia8}$~\cite{Sjostrand:2019zhc}, while the modeling of the detector response is accomplished through the $\texttt{Delphes}$ framework. To realistically reproduce the experimental environment, we employ the card with characteristics of the HL-LHC~\cite{DelphesHLLHCCard}.
\subsection{Production cross section}
\subsubsection{Signal}
\begin{figure*}[!htb]
    \centering
    \subfigure[]{\includegraphics[width=0.49\linewidth]{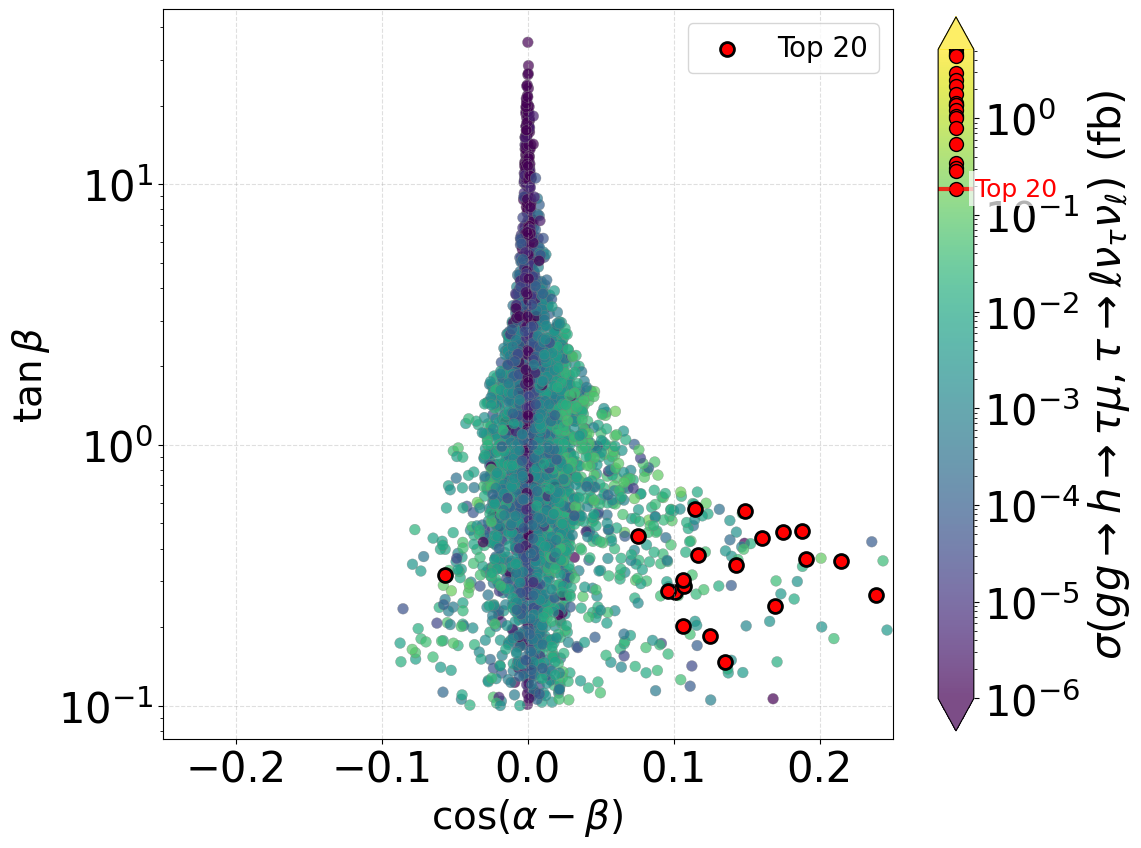}}
    \hfill
    \subfigure[]{\includegraphics[width=0.49\linewidth]{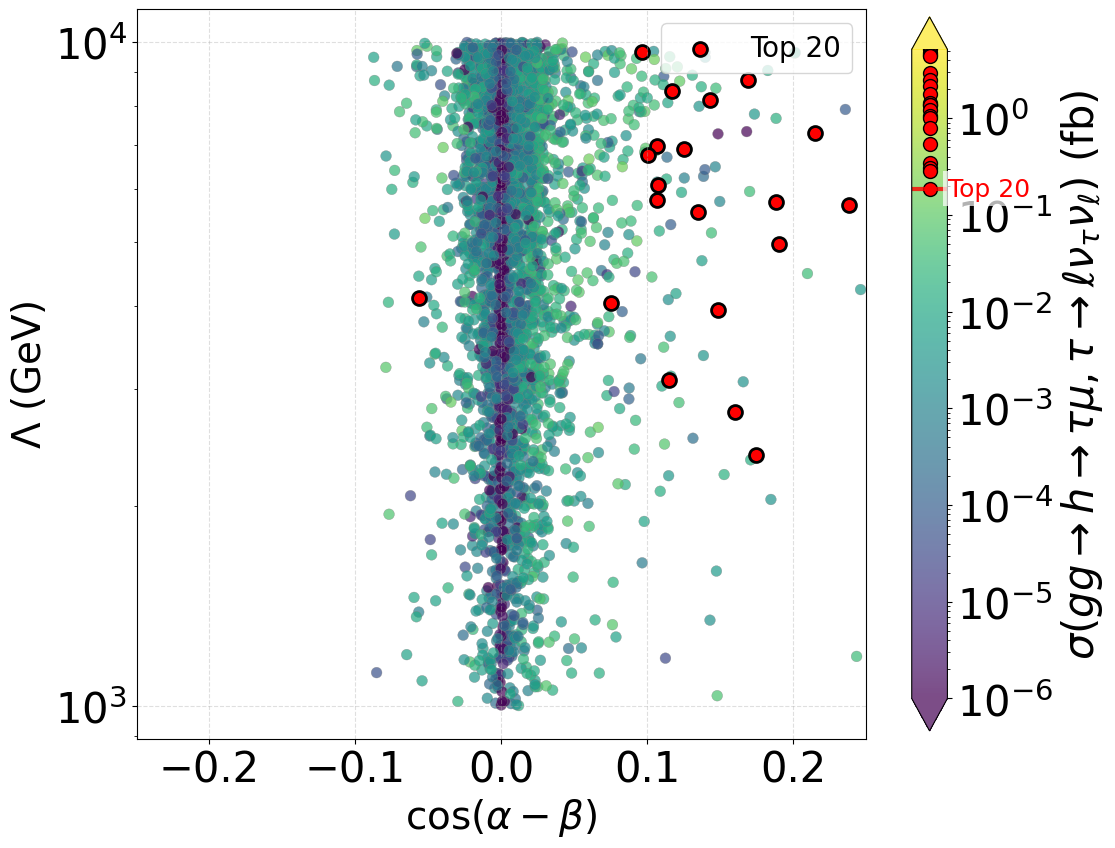}}
    \\
    \subfigure[]{\includegraphics[width=0.49\linewidth]{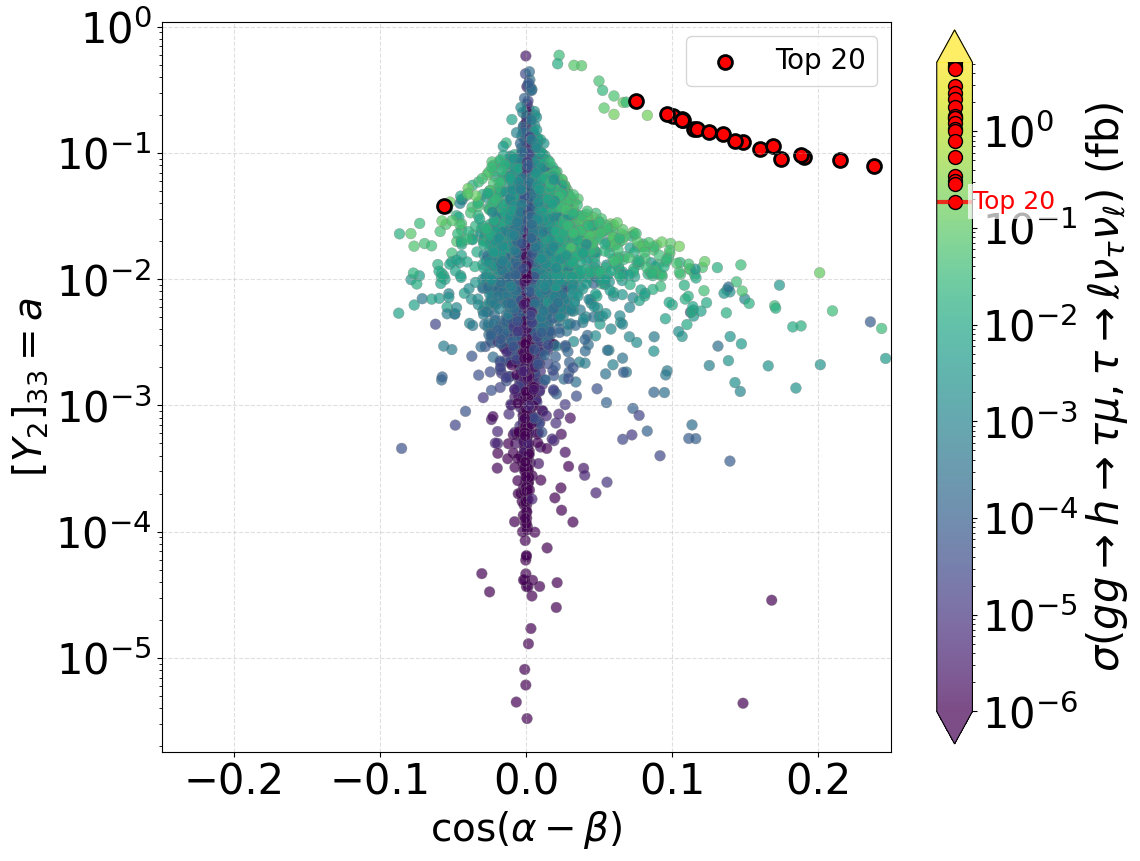}}
    \hfill
    \subfigure[]{\includegraphics[width=0.49\linewidth]{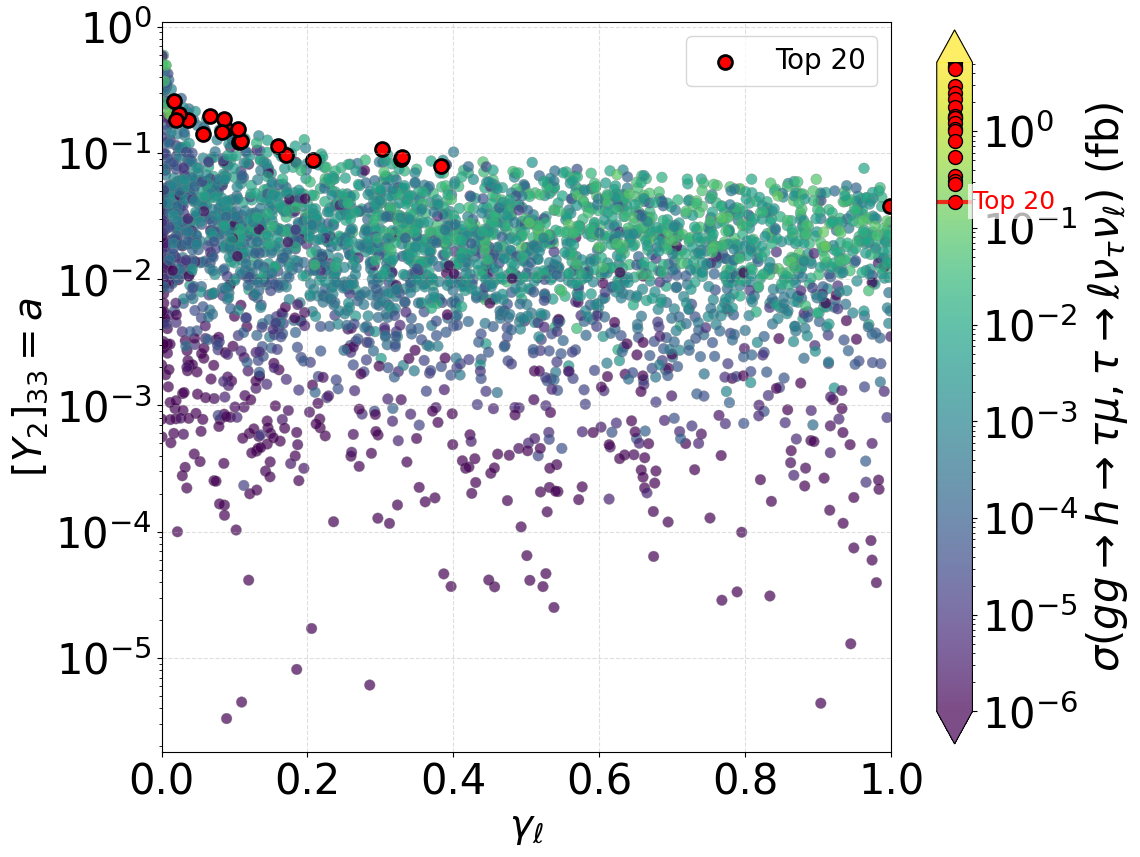}}
    \caption{Scatter plot of the cross section in the plane: $(a)$ $\cos(\alpha-\beta)$-$\tan\beta$, $(b)$ $\cos(\alpha-\beta)-\Lambda$, $(c)$ $\cos(\alpha-\beta)-a$ and  $(d)$ $\gamma_\ell-a$.}
    \label{Signal_XS}
\end{figure*}

We now scrutinize the behavior of the signal production cross-section across several parameter pairs. The data points displayed in the scatter plots of Fig.~\ref{Signal_XS} correspond exclusively to those that strictly satisfy the full set of constraints discussed in Sec.~\ref{sec:SecIII}.

The observed behavior of the cross section is due to the interaction $g_{h\tau\mu}$, which, upon omission of the mass term in Eq.~\eqref{yukawacoupling}, takes the form
\begin{equation}\label{ghtaumu}
    g_{h\tau\mu}=\frac{\cos(\alpha-\beta)}{\sqrt{2}}\sqrt{1+\tan^2\beta} \, \Bigg([\tilde{Y}_2]_{23}^{\ell}
+ \mathcal{F}^\ell_{23}\Bigg),
\end{equation}
where $[\tilde{Y}_2]_{23}^\ell$ and $\mathcal{F}^\ell_{23}$ are given as follows:

\begin{equation*}
[\tilde{Y}_2]_{23}^\ell = \frac{a\, m_\mu}{m_\tau-m_\mu} \sqrt{\gamma_\ell\Big((m_\tau/m_\mu)-\gamma_\ell-1\Big)},
\end{equation*}
and 
\begin{equation*}
\mathcal{F}^f_{23} = \frac{1}{2\Lambda}  v_{\eta_ 1} \,[\tilde{Y}_2]_{23}^\ell.
\label{Supresion_Lambda}
\end{equation*}
It is important to mention again that we consider case 15 for the present analysis, and it can change drastically if one considers another case of those presented in Sec.~\ref{sec:SecII}. 

From Eq.~\eqref{ghtaumu} we find a combination of parameters such that they increase the signal cross section: $\cos(\alpha-\beta)$, $\tan\beta$, $\gamma_\ell$ and $a$. The contribution of the term $\mathcal{F}^\ell_{ij}$ becomes suppressed by the cut-off scale $\Lambda$, which is of the order of TeV. As shown in Fig.~\ref{Signal_XS}(a)-(d), the production cross section of the signal $\sigma(gg\to h\to \tau\mu,\,\tau\to\ell\nu_\ell\nu_{\tau})$ ranges from $\mathcal{O}(10^{-6}-1)$ fb, where its largest values (top 20) are pointed by red circles enclosed in black outlines ---which are of order $10^{-1}-1$ fb---. Fig.~\ref{Signal_XS}$(a)$, $(b)$, $(c)$, $(d)$ corresponds to the plane $\cos(\alpha-\beta)$-$\tan\beta$, $\cos(\alpha-\beta)-\Lambda$, $\cos(\alpha-\beta)-a$ and $\gamma_\ell-a$, respectively.     
\subsubsection{Background}
The production cross sections at $\sqrt{s}=14$ TeV of the aforementioned background processes are given in Table~\ref{BGD_XSs}.


\begin{table}
\centering
\small
\caption{Main SM background processes for the signal $gg \to h \to \tau \mu \to \ell\mu + \mathrm{MET}$.}
\label{BGD_XSs}
\begin{tabular}{l c}
\toprule
\textbf{Process} & \textbf{Cross section (pb)} \\
\midrule
Drell-Yan  & 32.22  \\
\addlinespace
$t\bar{t}$   & 40.25\\
\addlinespace
$tW^\pm$  & 0.013\\
\addlinespace
$W^+W^-$  & 4.8 \\
\addlinespace
$ZZ$  & 0.43\\
\addlinespace
$W^\pm Z$& 0.55  \\
\bottomrule
\end{tabular}
\end{table}

\subsection{Multivariate Analysis}  
\begin{figure}[!htb]
		\centering
		\includegraphics[width=\linewidth]{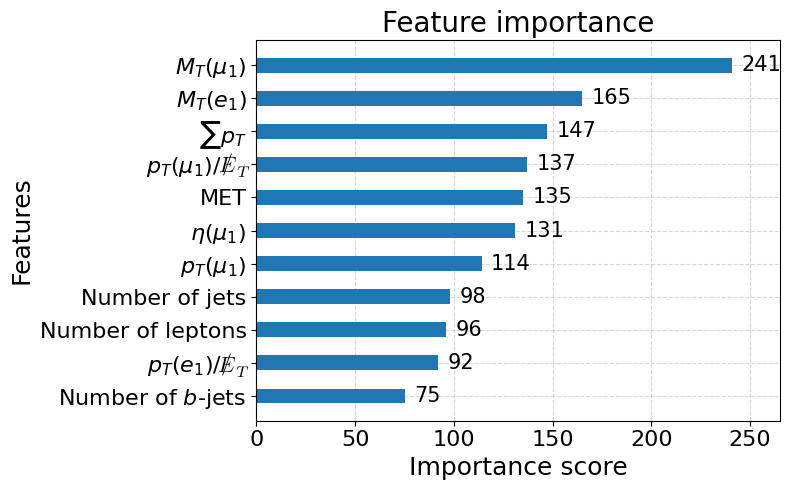}
		\caption{Ranking of the observables used for the BDT analysis. The index 1 in $e_1$ or $\mu_1$ stands for the electron or muon with the highest transverse momentum. }\label{features}
	\end{figure}

To enhance the signal-background discrimination, we adopt a multivariate analysis (MVA) selection that consolidates several kinematic observables into a single, denoted by XGB. The MVA training is performed using the Boosted Decision Trees (BDT) method~\cite{doi:10.1142/9789811234033_0002, Coadou_2022, Hastie:2009itz}, as implemented in the \texttt{XGBoost} library~\cite{Chen:2016:XST:2939672.2939785}, which provides a sophisticated gradient-boosting framework. The BDT classifiers are trained on the kinematic properties of the final-state particles, as presented in Fig. \ref{features}, which depicts the ranking of the observables used for the BDT analysis. Each of them are defined as follows:
\begin{itemize}
    \item $p_T(p_i)$: transverse momentum of the particle $p_i$,
    \item $\eta(p_i)$: pseudorapidity of the particle $p_i$,
    \item $\slashed{E}_T$: missing transverse energy (\texttt{MET}),
    \item $M_T^{\ell}=\sqrt{2p_T^{\ell}\slashed{E}_T(1-\cos(\Delta\phi))}$: transverse mass of lepton $\ell$ and missing energy,
    \item $\sum p_T$: sum of transverse momentum of the final visible particles.
\end{itemize}

\begin{figure*}[!htb]
    \centering
    \subfigure[]{\includegraphics[width=0.45\linewidth]{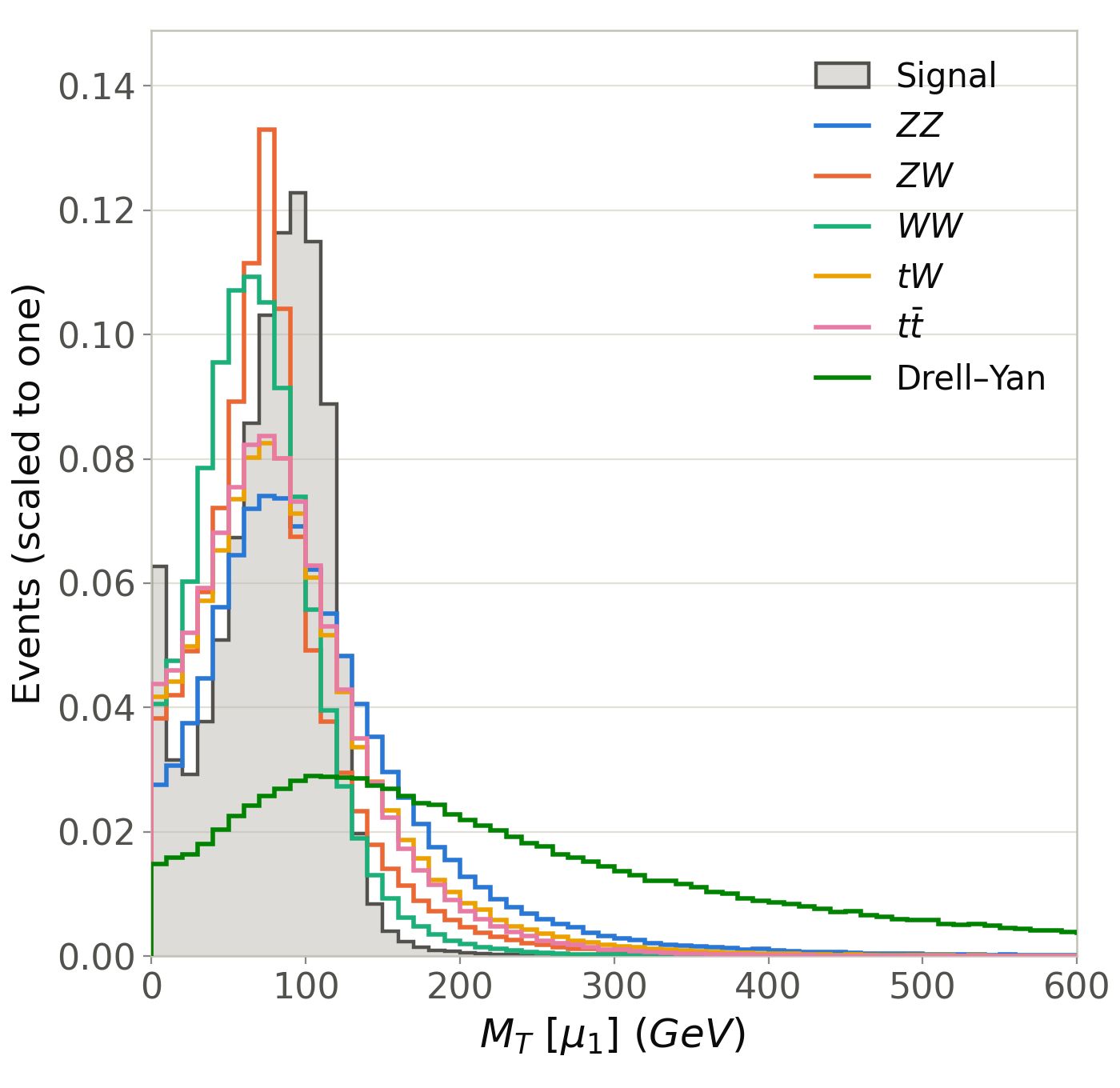}}
    \hfill
    \subfigure[]{\includegraphics[width=0.45\linewidth]{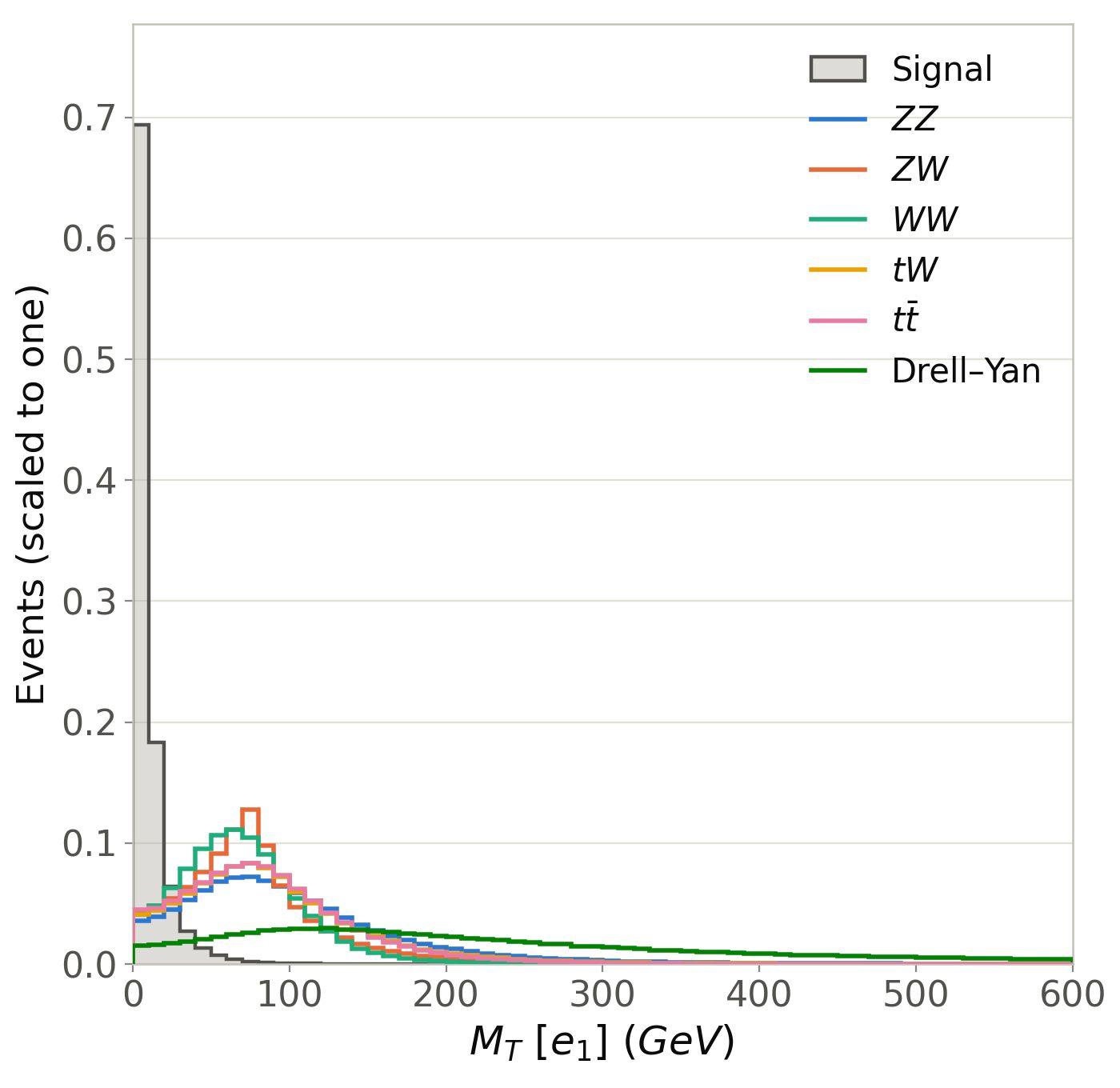}}
    \\
    \subfigure[]{\includegraphics[width=0.45\linewidth]{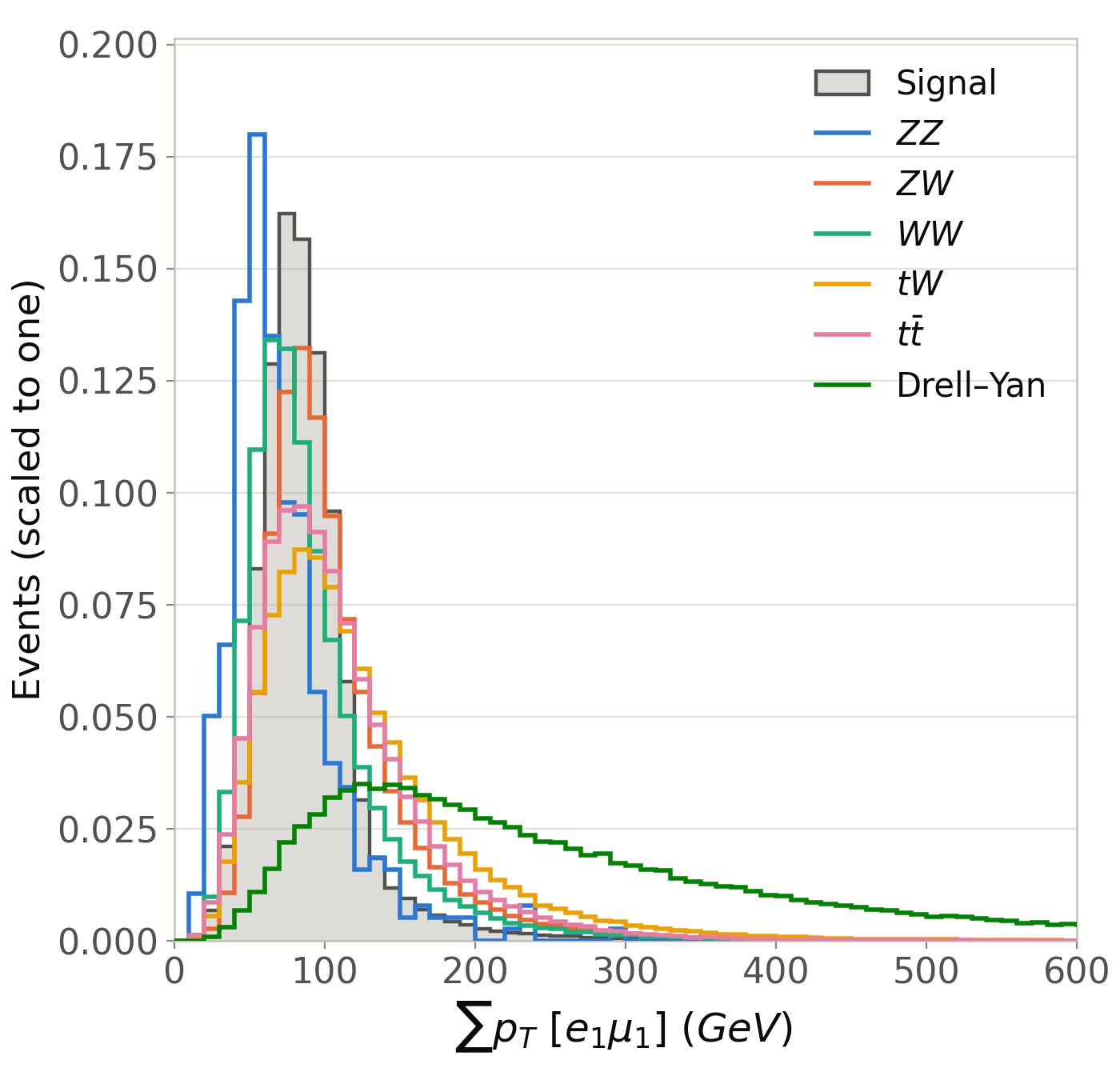}}
    \hfill
    \subfigure[]{\includegraphics[width=0.45\linewidth]{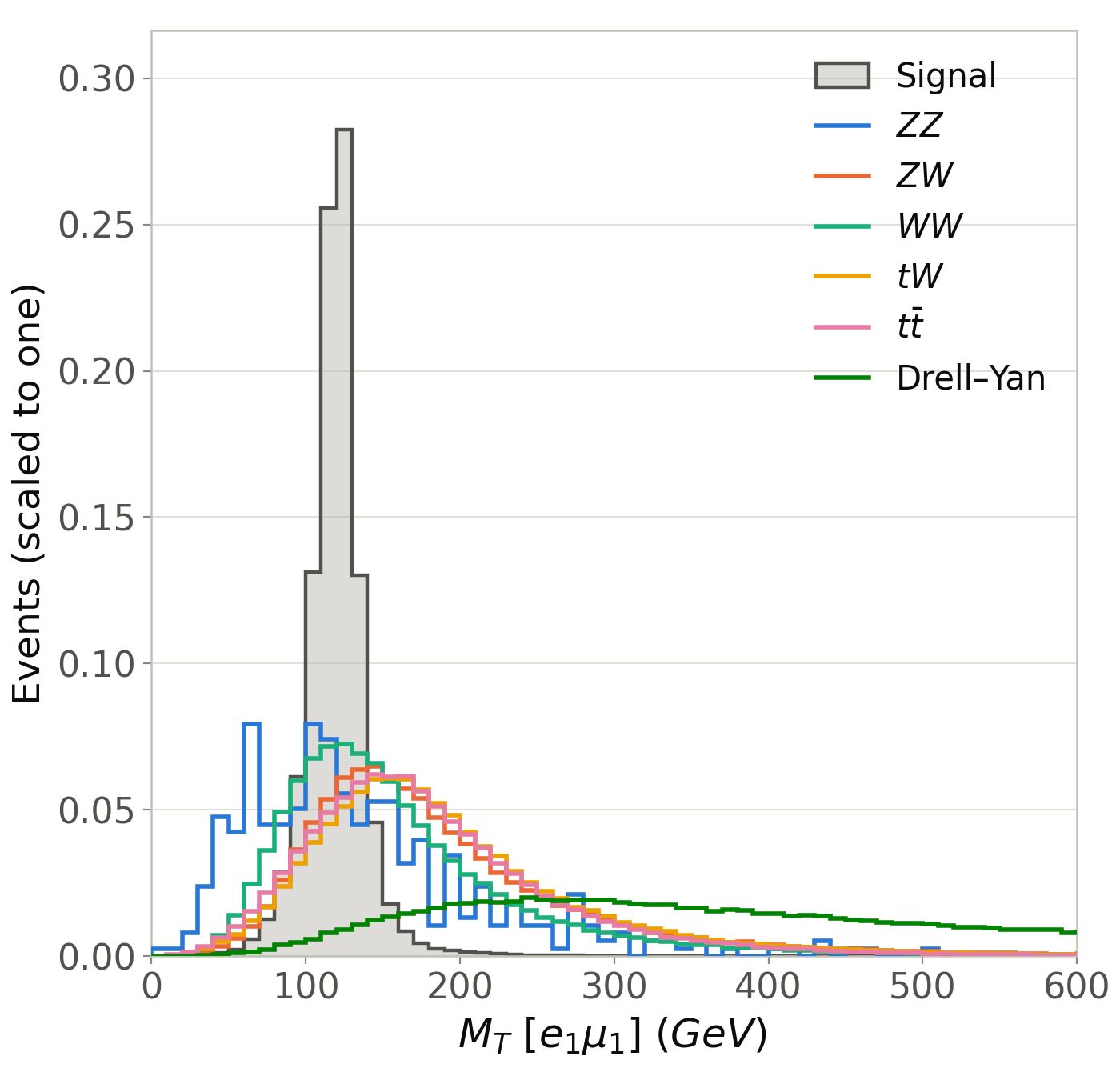}}
    \caption{Kinematic distributions of the signal and background processes: $(a)$ Transverse mass $M_T(\mu_1)$, $(b)$Transverse mass $M_T(e_1)$, $(c)$ Sum of the transverse momentum of the final visible particles $\sum p_T(e_1\mu_1)$ and $(d)$ Transverse mass of the final state $M_T(e_1\mu_1)$.}
    \label{distribuciones}
\end{figure*}

Training was performed using Monte Carlo simulated samples, which include the dominant irreducible backgrounds. The samples were generated with $1\times 10^6$ events for background and $5\times 10^5$ for signal, ensuring a balanced training set and robust statistical validation. The analysis revealed that most discriminating observables between the signal and the background are those shown in Fig.~\ref{distribuciones}(a)–(c). Meanwhile, Fig.~\ref{distribuciones}(d) shows the clearest observable to distinguish the signal: the transverse mass (centered around $m_h$), which includes missing energy attributed to undetected neutrinos.

\begin{figure}[!htb]
		\centering
		\includegraphics[width=0.9\linewidth]{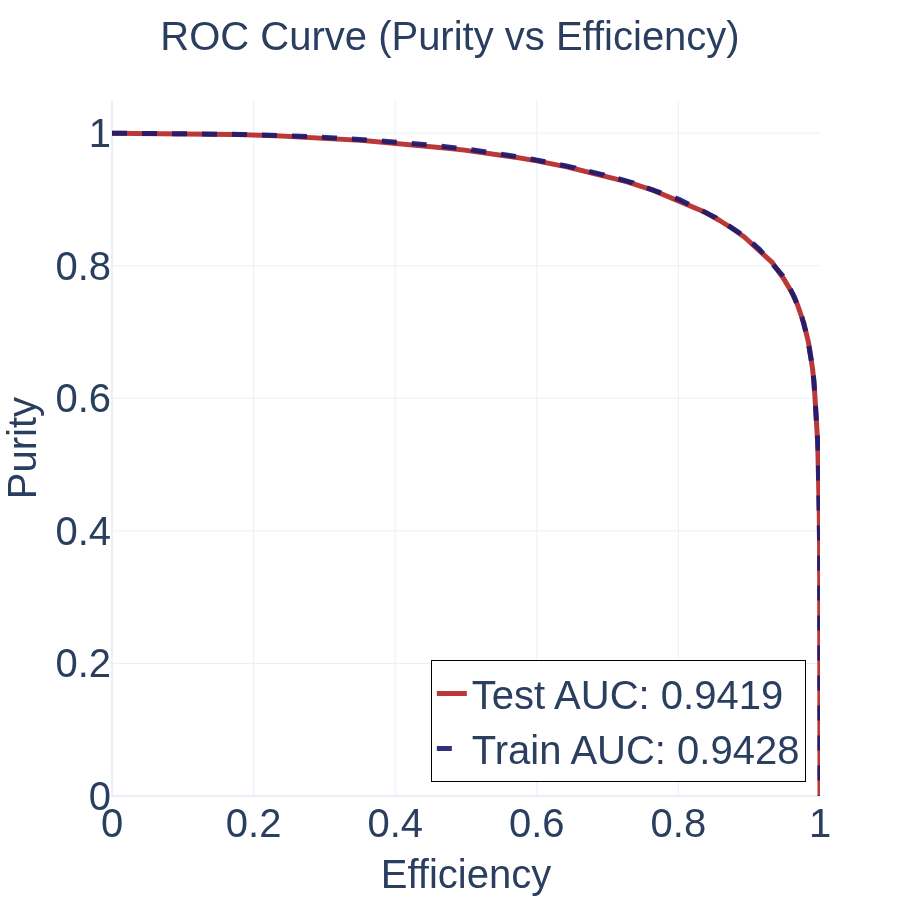}
		\caption{ROC curves (Purity vs. Efficiency) for the trained BDT classifier. }\label{fig:roc}
	\end{figure}

Figure~\ref{fig:roc} displays the Purity vs Efficiency curve, confirming a stable discriminatory power against the irreducible background. The final classifier achieves area-under-curve (AUC) values of $0.9428$ for training and $0.9419$ for testing.

\begin{figure}[!htb]
		\centering
		\includegraphics[width=\linewidth]{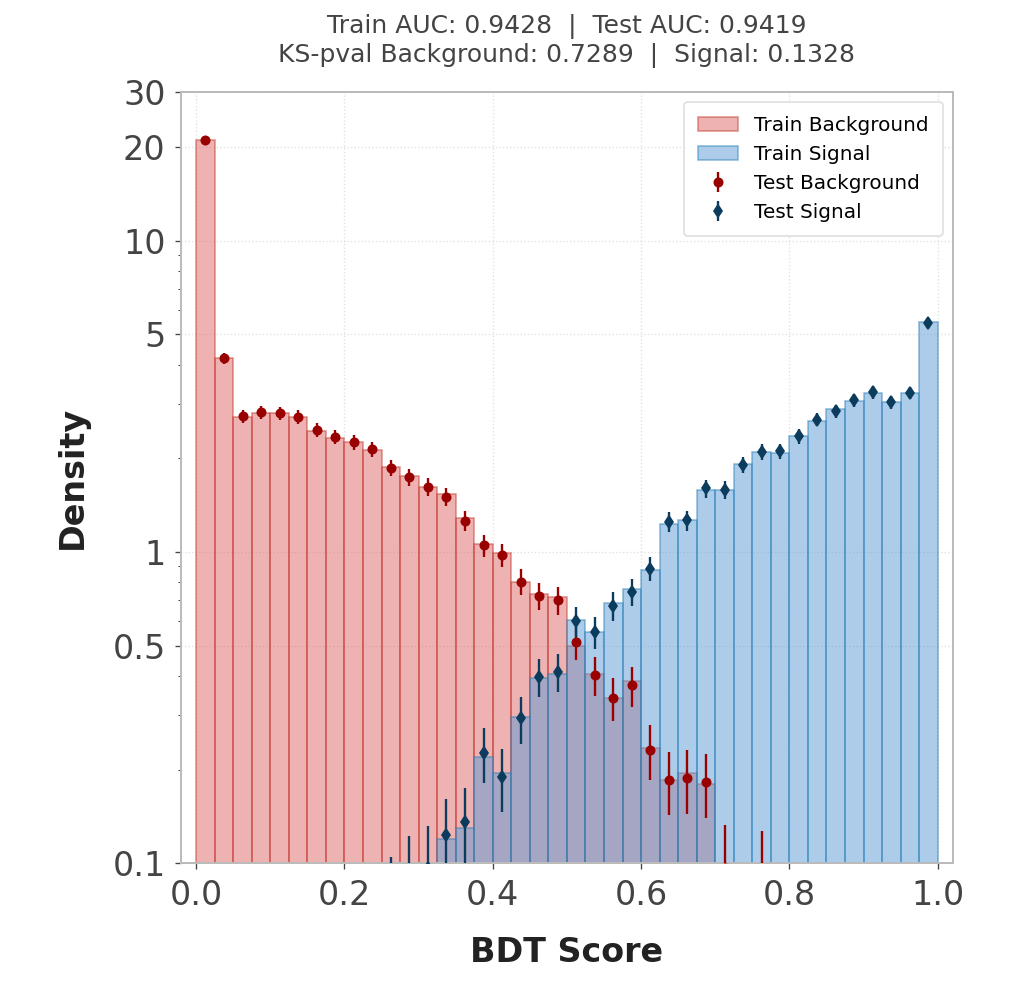}
		\caption{ Distribution of BDT scores for signal and background events across training and testing
samples.}\label{fig:bdt_output}
	\end{figure}

Meanwhile, in Fig.~\ref{fig:bdt_output} the corresponding BDT output distributions are shown, revealing a marked concordance between the training and testing samples. This agreement is quantitatively validated by the Kolmogorov--Smirnov $p$-values (KS-$p$), which robustly rule out overtraining, yielding values of $0.7289$ for the background class and $0.1328$ for the signal class, which reside well within conventionally accepted thresholds. 

Signal and background samples were scaled to the expected event yields, calculated from the integrated luminosity and corresponding cross sections. Then, we calculate the figure of merit, the signal significance, defined as $\mathcal{N}_S/\sqrt{\mathcal{N}_S+\mathcal{N}_B}$, where $\mathcal{N}_S$ and $\mathcal{N}_B$ are the expected numbers of signal and background events. With the goal of maximizing the significance, the BDT hyperparameters were optimized using the Optuna framework \cite{akiba2019optuna}\footnote{To facilitate the reproduction of the results reported herein, the hyperparameters employed in the BDT analysis as well as the datasets are provided upon request.}. Furthermore, we implement two criteria on the final state: a \mbox{$b$-jet} veto and the requirement of exactly two leptons of opposite charge. These conditions successfully enhance the signal by up to 10\%.

\subsection{Signal significance}
\begin{figure*}[!htb]
    \centering
    \subfigure[]{\includegraphics[width=0.49\linewidth]{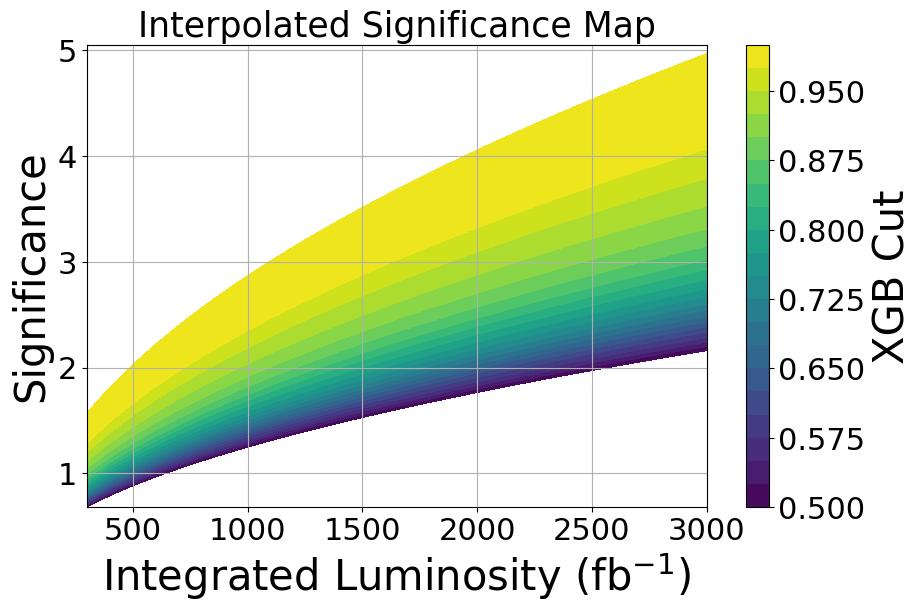}}
    \hfill
    \subfigure[]{\includegraphics[width=0.49\linewidth]{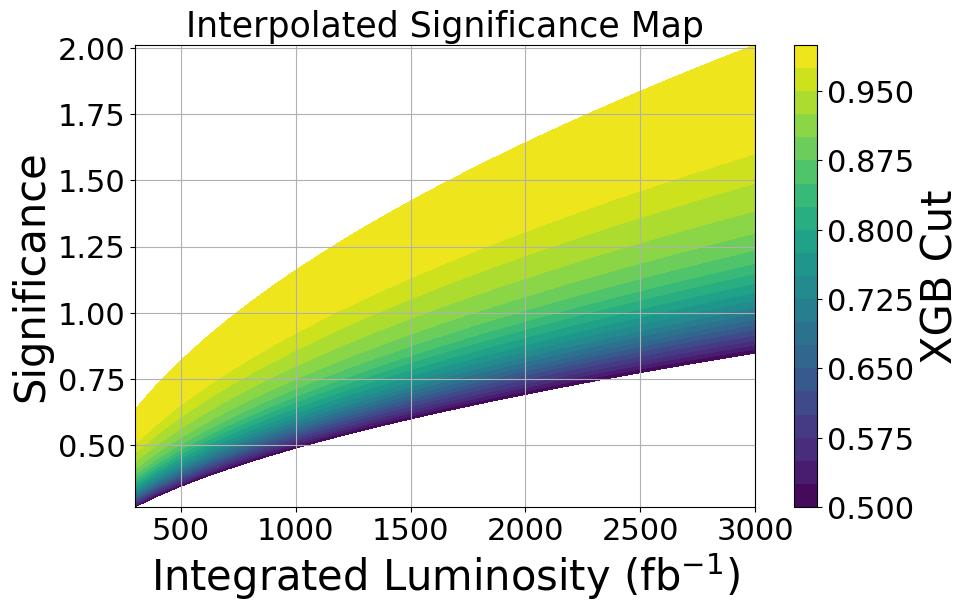}}
    \caption{Signal significance as a function of the integrated luminosity and the XGB cut corresponding to $(a)$ $\sigma_1=5.1$ fb, $(b)$ $\sigma_2=2$ fb.}
    \label{Significances}
\end{figure*}

We now turn to present our most relevant results for the phenomenological analysis. Figure \ref{Significances} shows the signal significance as a function of the integrated luminosity and the XGB cut for two illustrative values of the cross section $\sigma(gg\to h\to \tau\mu,\,\tau\to\ell\nu_\ell\nu_{\tau})$, denoted $\sigma_1=5.1$ fb and $\sigma_2=2$ fb, both allowed for all the constraints studied in Sec.~\ref{sec:SecIII} and reported in Fig.~\ref{Signal_XS}.

Fig.~\ref{Significances}(a) is the most favorable scenario we found ($\sigma_1=5.1$ fb), which, as expected, predicts the highest significance, $\approx 5\sigma$, once the integrated luminosity reaches 3000 fb$^{-1}$ and using a XGB of 0.095, suggesting a possible  discovery of the signature proposed in this work. For luminosities in the $(1000,\,3000)$ fb$^{-1}$ interval corresponding to XGB cuts between (0.95, 0.75), respectively, we predict $3\sigma$ evidence for the signature.
Figure~\ref{Significances}(b) shows that limits can be imposed in $2\sigma$ on a cross section of $\sigma(gg\to h\to\tau\mu\to \ell\nu\nu\mu)=2$ fb.  

This analysis could be enhanced by evaluating additional cases among the 14 previously outlined, as they incorporate supplementary parameters capable of improve the signal. While Case 15 serves as the most straightforward configuration—utilized here primarily for illustrative purposes and the phenomenological validation of the proposed model—the remaining cases offer deeper insights.

\section{Conclusions}\label{sec:SecV}

In this work, we have presented the theoretical construction and phenomenological validation of an extended Texturized Two-Higgs Doublet Model (2HDM-Tx) invariant under a discrete, non-abelian $D_4$ dihedral symmetry. The core structural virtue of this framework lies in its ability to provide a rigorous symmetry-based justification for the standard four-zero texture in the fermion mass matrices, thus translating empirical texture ans\"atze into a fundamental structural property of the Yukawa Lagrangian and the scalar potential.

The model architecture is built upon the unique faithful representation assignment of the fields under the order-8 $D_4$ group. By assigning first generation fermions to the $\mathbf{1_{++}}$ singlet representation and the heavier generations to the $\mathbf{2}$ doublet representation, renormalizable Yukawa couplings directly involving the first generation are strictly forbidden by the $D_4$ multiplication rules. This automatically establishes structural zeros at the $(1,1)$, $(1,3)$, and $(3,1)$ entries of both $Y_1$ and $Y_2$ in the gauge basis. To generate the remaining four parameters ($a, b, c, d$), the scalar sector is non-trivially extended by a complex flavon doublet $\eta \sim \mathbf{2}$ alongside the two Higgs doublets $\Phi_1 \sim \mathbf{1_{+-}}$ and $\Phi_2 \sim \mathbf{1_{-+}}$. After the flavon doublet acquires a vacuum expectation value $\langle\eta\rangle = (v_{\eta_1}, v_{\eta_2})^T \neq 0$, effective renormalizable terms are generated via dimension-5 operators of the form $\frac{y}{\Lambda}(\overline{Q}_L \eta)_{\text{inv}}\Phi_k \psi_R$. This multi-scale symmetry-breaking mechanism naturally provides a structural rationale for the fermion mass hierarchy, where the smallness of the first-generation masses is naturally suppressed by the heavy UV cutoff scale $\Lambda$. Furthermore, the explicit proof of $D_4$ invariance confirms that the full renormalizable scalar potential, $V_{D_4}$, remains stable and successfully isolates the $D_4$-singlet quadratic and quartic operators. Additionally, while the model inherently predicts a viable scalar singlet dark matter candidate, a rigorous analysis of its cosmological relic abundance and direct detection constraints is deferred to a future standalone study.

The structural viability of this algebraic construction has been rigorously validated through its phenomenological implications. A systematic scan over the parameter space shows that the model successfully accommodates current LHC Higgs data and satisfies the stringent constraints from tree-level flavor-changing neutral currents (FCNCs) and lepton flavor violation (LFV) processes. Remarkably, the mixing induced by the flavon components within the CP-even and CP-odd scalar sectors provides a simultaneous explanation for the persistent neutral resonance excesses reported at $95\text{ GeV}$ and $152\text{ GeV}$.

Finally, to probe the distinctive signature of this flavor framework, we performed a Monte Carlo simulation of the LFV process $h \rightarrow \tau \mu$ at the HL-LHC ($\sqrt{s} = 14\text{ TeV}$). By implementing a multivariate analysis using Boosted Decision Trees (BDT) via the XGBoost library, we demonstrated that the signal can be efficiently isolated from the SM background. For the predictive benchmark of Case 15---where $Y_2$ is strictly sparse with a single non-vanishing diagonal entry---a formal $5\sigma$ discovery significance is attainable at a total integrated luminosity of $3000\text{ fb}^{-1}$. Coupled with a stable scalar singlet $\chi \sim \mathbf{1_{++}}$ that serves as a viable dark matter candidate, the 2HDM-Tx stands as a highly predictive, structurally sound, and testable extension of the SM.


\section*{Acknowledgments}
The work of M. A. Arroyo-Ure\~na and H. Hernández-Arellano is supported by “Estancias posdoctorales por México (SECIHTI)”, “Sistema Nacional de Investigadoras e Investigadores” (SNII), M.A.A.U also thanks the support of the SECIHTI project No. CBF-2025-G-1187. E. D\'iaz would like to acknowledge the Aspen Center for Physics (ACP), where this work was performed in part.
\newpage
\section{Appendix}
\subsection{Proof of \(D_4\) invariance for case 15}~\label{InvCases}

\subsubsection{Case 15}

\begin{equation}
Y_1 =
\begin{pmatrix}
0 & d & 0\\
d^* & c & b\\
0 & b^* & 0
\end{pmatrix},
\qquad
Y_2 =
\begin{pmatrix}
0 & 0 & 0\\
0 & 0 & 0\\
0 & 0 & a
\end{pmatrix}.
\end{equation}

$\Phi_1$ generates $d,c,b$; \quad $\Phi_2$ generates $a$.

\medskip

\textbf{Term $d$ in $Y_1$}: $\bar{Q}_L^2\Phi_1u_R^1$ (position $(2,1)$ of $Y_1$)

1. Under $r$:
$\bar{Q}_L^2 \to \bar{Q}_L^2$, $\Phi_1 \to +\Phi_1$, $u_R^1 \to +u_R^1$.
Total: $(+1)\times(+1)\times(+1) = +1$.

2. Under $s$:
$\bar{Q}_L^2 \to \bar{Q}_L^2$, $\Phi_1 \to -\Phi_1$, $u_R^1 \to +u_R^1$.
The Clebsch-Gordan coefficient of the $2\times 1_{+-}\times 1_{++}$ contraction exactly cancels the minus sign from $\Phi_1$. Total: $+1$.

\medskip

\textbf{Term $c$ in $Y_1$}: $\bar{Q}_L^2\Phi_1u_R^2$ (position $(2,2)$ of $Y_1$)

1. Under $r$:
$\bar{Q}_L^2 \to \bar{Q}_L^2$, $\Phi_1 \to +\Phi_1$, $u_R^2 \to -u_R^3$ (first component of the right doublet).
The $2\otimes 2$ singlet contraction with $\Phi_1$ (the $\eta_1\eta_2 + \eta_2\eta_1$ channel) produces an overall $+1$ after summing both components.

2. Under $s$:
$\bar{Q}_L^2 \to \bar{Q}_L^2$, $\Phi_1 \to -\Phi_1$, $u_R^2 \to +u_R^2$.
The CG coefficient compensates the minus sign, total $+1$.

\medskip

\textbf{Term $b$ in $Y_1$}: $\bar{Q}_L^2\Phi_1u_R^3$ (position $(2,3)$ of $Y_1$)

1. Under $r$:
$\bar{Q}_L^2 \to \bar{Q}_L^2$, $\Phi_1 \to +\Phi_1$, $u_R^3 \to +u_R^2$ (second component under $r$).
The $2\otimes 2$ singlet contraction yields total $+1$.

2. Under $s$:
$\bar{Q}_L^2 \to \bar{Q}_L^2$, $\Phi_1 \to -\Phi_1$, $u_R^3 \to -u_R^3$.
The CG cancels the signs, total $+1$.

\medskip

\textbf{Term $a$ in $Y_2$}: $\bar{Q}_L^3\Phi_2u_R^3$ (position $(3,3)$ of $Y_2$)

1. Under $r$:
$\bar{Q}_L^3 \to \bar{Q}_L^3$, $\Phi_2 \to -\Phi_2$, $u_R^3 \to +u_R^2$.
The $2\otimes 2$ singlet contraction with $\Phi_2$ gives total $+1$.

2. Under $s$:
$\bar{Q}_L^3 \to -\bar{Q}_L^3$, $\Phi_2 \to +\Phi_2$, $u_R^3 \to -u_R^3$.
Two minus signs from the doublets cancel, giving total $+1$.

\medskip

All other entries are zero and possess a non-trivial total representation (explicit phase $\neq +1$). Thus, in Case 15 the non-zero entries are precisely the $D_4$-singlet operators and the Lagrangian is invariant.

\subsection{Proof of \(D_4\) invariance of the scalar potential and Yukawa Lagrangian}~\label{InvPotential}

We use the generator actions derived in the previous subsection:
\begin{align}
r\cdot\Phi_1&=+\Phi_1, && s\cdot\Phi_1=-\Phi_1, \label{eq:phi1-repeat} \\
r\cdot\Phi_2&=-\Phi_2, && s\cdot\Phi_2=+\Phi_2, \\
r\cdot\eta&=\begin{pmatrix}0&-1\\1&0\end{pmatrix}\eta, &&
s\cdot\eta=\begin{pmatrix}1&0\\0&-1\end{pmatrix}\eta.
\end{align}

\subsubsection{Scalar potential \(V_{D_4}\)}
The full renormalisable potential is
\begin{align}\label{ScalarPotential}
V_{D_4} &= m_{11}^2 |\Phi_1|^2 + m_{22}^2 |\Phi_2|^2 
+ \frac{\lambda_1}{2}(|\Phi_1|^2)^2 + \frac{\lambda_2}{2}(|\Phi_2|^2)^2\nonumber\\ 
&+ \lambda_3 |\Phi_1|^2|\Phi_2|^2 + \lambda_4 |\Phi_1^\dagger\Phi_2|^2 \nonumber 
+ m_\eta^2 |\eta|^2 + \lambda_\eta (|\eta|^2)^2 \\
&+ \lambda_{\Phi\eta} \bigl(|\Phi_1|^2 + |\Phi_2|^2\bigr)|\eta|^2 
+ \kappa_1 \bigl(\Phi_1^\dagger \Phi_2 \eta_1\eta_2 + \text{h.c.}\bigr)\nonumber\\
&+ \kappa_2 \bigl(\Phi_1^\dagger \Phi_1 \eta_1^2 + \Phi_2^\dagger \Phi_2 \eta_2^2 + \text{h.c.}\bigr)+ V_{\text{DM}}, 
\end{align}

All quadratic and quartic terms are \(D_4\)-singlets because they are built from \(D_4\)-invariant bilinears. Explicit verification under generators:
\\
\\
- Mass terms \(|\Phi_1|^2\), \(|\Phi_2|^2\), \(|\eta|^2\):  
  \(r:\) each factor acquires phase \(+1\) (or \(-1\) for \(\Phi_2\) under \(r\), but squared \(\to+1\)).  
  \(s:\) \(\Phi_1\to-\Phi_1\) \(\Rightarrow\) \(|\Phi_1|^2\to|\Phi_1|^2\); \(\Phi_2\to+\Phi_2\); \(\eta\to(\eta_2,-\eta_1)\) \(\Rightarrow\) \(|\eta|^2\to|\eta|^2\).
\\
\\
- Quartic terms without \(\kappa\):  
  \(|\Phi_1|^4\), \(|\Phi_2|^4\), \(|\Phi_1|^2|\Phi_2|^2\), \(|\Phi_1^\dagger\Phi_2|^2\), \(|\eta|^4\), and \(|\Phi_i|^2|\eta|^2\) are products of the above invariants, hence unchanged under both \(r\) and \(s\).
\\
\\
- \(\kappa_1\) term \(\Phi_1^\dagger\Phi_2\eta_1\eta_2+\text{h.c.}\):  
  Under \(r\): \(\Phi_1^\dagger\to+\Phi_1^\dagger\), \(\Phi_2\to-\Phi_2\), \(\eta_1\to-\eta_2\), \(\eta_2\to\eta_1\) \(\Rightarrow\) \(\eta_1\eta_2\to(-\eta_2)\eta_1=-\eta_1\eta_2\).  
  Overall phase: \((+1)\times(-1)\times(-1)=+1\).  
  Under \(s\): \(\Phi_1^\dagger\to-\Phi_1^\dagger\), \(\Phi_2\to+\Phi_2\), \(\eta_1\to\eta_1\), \(\eta_2\to-\eta_2\) \(\Rightarrow\) \(\eta_1\eta_2\to\eta_1(-\eta_2)=-\eta_1\eta_2\).  
  Overall phase: \((-1)\times(+1)\times(-1)=+1\).  
  Hermitian conjugate transforms identically.
\\
\\
- \(\kappa_2\) term \(\Phi_1^\dagger\Phi_1\eta_1^2+\Phi_2^\dagger\Phi_2\eta_2^2+\text{h.c.}\):  
  Under \(r\): \(\Phi_1^\dagger\Phi_1\to|\Phi_1|^2\), \(\eta_1^2\to(-\eta_2)^2=\eta_2^2\); \(\Phi_2^\dagger\Phi_2\to|\Phi_2|^2\), \(\eta_2^2\to\eta_1^2\).  
  The sum \(\eta_1^2+\eta_2^2\) is exchanged but the coefficients are identical, so invariant.  
  Under \(s\): \(\eta_1^2\to\eta_1^2\), \(\eta_2^2\to\eta_2^2\) (signs square to \(+1\)); \(\Phi\) bilinear invariants.  
  Hermitian conjugate unchanged.

Thus \(V_{D_4}\) is fully \(D_4\)-invariant.
\\
\\
\subsubsection{Yukawa Lagrangian \(\mathcal{L}_Y\)}
For the up-quark sector (identical for \(d,e,\nu\)):
\begin{eqnarray}
    \mathcal{L}_Y^U &=& Y_d^U\,\bar{Q}_L^2\Phi_1 u_R^1
+ Y_c^u\,\bar{Q}_L^2\Phi_2 u_R^2
+ Y_b^u\,\bar{Q}_L^2\Phi_2 u_R^3\nonumber\\
&+& Y_a^u\,\bar{Q}_L^3\Phi_2 u_R^3
+ \frac{y_\eta^u}{\Lambda}(\bar{Q}_L\eta)_{\rm inv}\Phi_2 u_R + \text{h.c.},
\end{eqnarray}

where \((\bar{Q}_L\eta)_{\rm inv}\) is the unique \(D_4\)-singlet contraction of two doublets.

- Each direct term \(\bar{\psi}_L\Phi_k\psi_R\) transforms with the phase product of the three fields. By construction of the representation assignment, every allowed term is a singlet while forbidden positions (including \((1,1)\), \((1,3)\), \((3,1)\)) have non-zero total phase under \(r\) or \(s\).

- Effective term \((\bar{Q}_L\eta)_{\rm inv}\Phi_2 u_R\):  
  The contraction \((\bar{Q}_L\eta)_{\rm inv}\) is the singlet component of \(\mathbf{2}\otimes\mathbf{2}\to\mathbf{1}_{++}\). Under \(r\) and \(s\) it picks phase \(+1\) (by definition of the CG). \(\Phi_2\) and \(u_R\) (both \(\mathbf{2}\)) combine with it to give total singlet.

Since the fermion assignments (\(Q_L^1,u_R^1\sim\mathbf{1}_{++}\), \((Q_L^{2,3},u_R^{2,3})\sim\mathbf{2}\)) are identical across sectors, the full \(\mathcal{L}_Y=\mathcal{L}_Y^u+\mathcal{L}_Y^d+\mathcal{L}_Y^e+\mathcal{L}_Y^\nu\) is \(D_4\)-invariant. After EWSB and \(\langle\eta\rangle\neq0\), the effective Yukawa matrices \(Y_1^f\) and \(Y_2^f\) inherit the same invariance properties.

This completes the proof that both \(V_{D_4}\) and \(\mathcal{L}_Y\) are invariant under the full \(D_4\) group.
\bibliography{biblio}
\end{document}